\documentclass[usenatbib]{mnras}
\usepackage[T1]{fontenc}
\usepackage[utf8]{inputenc}
\usepackage{blindtext}
\usepackage{comment}
\usepackage{amsmath}
\usepackage[dvipsnames]{xcolor}
\usepackage{array}
\usepackage{graphicx}
\usepackage{tikz}
\graphicspath{ {./} }
\newcolumntype{P}[1]{>{\centering\arraybackslash}p{#1}}

\title[A tell-tale signature of merging MBHBs]{Disappearing thermal X-ray emission as a tell-tale signature of merging massive black hole binaries}

\author[Krauth et al.]{ Luke Major Krauth$^{1}$\thanks{E-mail: LMK2202@columbia.edu},
Jordy Davelaar$^{2,3,4}$,
Zoltán Haiman$^{1,2}$, \and
John Ryan Westernacher-Schneider$^{5}$,
Jonathan Zrake$^{6}$,
and Andrew MacFadyen$^{7}$
\\
$^{1}$Department of Physics, Columbia University, New York, NY 10027, USA\\
$^{2}$Department of Astronomy, Columbia University, New York, NY 10027, USA\\
$^{3}$Astrophysics Laboratory, Columbia University, 550 W 120th St, New York, NY 10027, USA\\
$^{4}$Center for Computational Astrophysics, Flatiron Institute, 162 Fifth Avenue, New York, NY 10010, USA\\
$^{5}$Leiden Observatory, Leiden University, P.O. Box 9513, 2300 RA Leiden, The Netherlands\\
$^{6}$Department of Physics and Astronomy, Clemson University, Clemson, SC 29634, USA\\
$^{7}$Center for Cosmology and Particle Physics, Physics Department, New York University, New York, NY 10003, USA\\
}

\date{Accepted 2023 October 2. Received 2023 September 29; in original form 2023 April 5}
\pubyear{2023}

\begin{document}
\maketitle

\begin{abstract}
\normalsize
The upcoming Laser Interferometer Space Antenna (LISA) is expected to detect gravitational waves (GWs) from massive black hole binaries (MBHB). Finding the electromagnetic (EM) counterparts for these GW events will be crucial for understanding how and where MBHBs merge, measuring their redshifts, constraining the Hubble constant and the graviton mass, and for other novel science applications. However, due to poor GW sky localisation, multi-wavelength, time-dependent electromagnetic (EM) models are needed to identify the right host galaxy. We studied merging MBHBs embedded in a circumbinary disc using high-resolution two-dimensional simulations, with a $\Gamma$-law equation of state, incorporating viscous heating, shock heating, and radiative cooling.
We simulate the binary from large separation until after merger, allowing us to model the decoupling of the binary from the circumbinary disc (CBD). We compute the EM signatures and identify distinct features before, during, and after the merger. Our main result is a multi-band EM signature: we find that the MBHB produces strong thermal X-ray emission until 1-2 days prior to the merger. However, as the binary decouples from the CBD, the X-ray-bright minidiscs rapidly shrink in size, become disrupted, and the accretion rate drops precipitously.
As a result, the thermal X-ray luminosity drops by orders of magnitude, and the source remains X-ray dark for several days, regardless of any post-merger effects such as GW recoil or mass loss. Looking for the abrupt spectral change where the thermal X-ray disappears is a tell-tale EM signature of LISA mergers that does not require extensive pre-merger monitoring.
\end{abstract}


\section{Introduction}

Over the past decade, significant progress has been made in the detection of gravitational waves (GWs) by the Laser Interferometer Gravitational-Wave Observatory \citep{aLIGO2015}, Virgo \citep{acernese2014} and the Kamioka Gravitational Wave Detector (KAGRA; \citealt{akutsu2021}). These detectors are sensitive to GWs with frequencies between $\sim$10 Hz and  $\sim$1 kHz, and have detected nearly 100 mergers involving stellar-mass black holes (BHs) and neutron stars. The Laser Interferometer Space Antenna (LISA; \citealt{amaro-seoane2017}) will access new frequencies between $\sim$0.1~mHz and $\sim$1~Hz, opening a new window to detect binaries with much larger masses and wider orbital separations.   

While detecting these new sources and characterising their GW signatures will be impactful by itself, it is recognised that the combined detection of GW and electromagnetic (EM) signatures will enable a range of additional science.  This so-called "multi-messenger" combination can provide novel probes of alternative theories of gravity \citep[such as those requiring extra dimensions, see][]{rham2018}, the mass of the graviton \citep[such as in recent ghost-free massive gravity models, see][]{hassan2012}, and the distance-redshift relation \citep[including measurements of the Hubble constant H$_0$, see][]{schutz1986}.
On the astrophysics side, multi-messenger detections will yield novel constraints on accretion physics and the co-evolution of galaxies and their central massive BHs (see, e.g., \citealt{baker2019} for a brief review and references).

Possible EM signatures from merging MBHBs can be divided into two broad categories: pre- and post-merger~\citep[see, e.g.][for a comprehensive review]{bogdanovic2022}.
Before their merger, MBHBs are expected to be typically surrounded by a circumbinary disc (CBD), delivered to the nucleus by the preceding galaxy merger~\citep{begelman1980}. In recent years, a consensus has been reached by hydrodynamical simulations describing a MBHB+CBD system: while the binary creates a low-density central cavity, out to a few times its separation, the individual components are fed efficiently through narrow streams, fuelling individual "minidiscs". Pre-merger EM signatures can therefore originate from one of several locations: the CBD, the minidiscs, or the streams, either colliding with each other or the wall of the cavity~\citep[see, e.g.][and references therein]{westernacher-schneider2022}. 

Early work suggested that because the GW-driven inspiral time ($t_{\rm gw} \propto a^4$) decreases more rapidly than the viscous inflow time (e.g. $t_{\rm visc} \propto a^{7/5}$; \citealt{pringle1991}), the binary would outrun the circumbinary disc~\citep{liu2003,milosavljevic2005}, starving and dimming the binary until the cavity re-fills in approximately a viscous time after the merger (which would take at least years or decades). However, long-term 2D hydrodynamical simulations following the GW-driven inspiral did not reveal evidence for such a clear "decoupling". Instead, they showed that the circumbinary gas could follow the inspiralling binary and feed the individual BHs all the way until the merger~\citep{farris2015a,tang2018}.  Relativistic simulations of fewer orbits closer to the merger also suggest that efficient fuelling of the minidiscs at these late states is possible, provided circumbinary gas still surrounds the binary~\citep{noble2012,bowen2017}.

Pre-merger EM signatures include periodic brightness fluctuations on approximately the binary's orbital timescale, driven by hydrodynamic modulations~\citep[see, e.g.][for early works]{artymowicz1996,hayasaki2007,macfadyen2008,Roedig2011,shi2012,noble2012,dorazio2013,farris2014} and/or kinematic Doppler~\citep{bode2010,dorazio2015,haiman2017} and lensing~\citep{dorazio2018,davelaar2022,davelaar2022b} effects. The buildup of a non-axisymmetric density distribution near the edge of the cavity, known as a "lump", modulates accretion onto the minidiscs, causing periodicity on several times the orbital timescale, as well as on beat periods between the binary and the lump~\citep[e.g.][]{shi2012,dorazio2013,farris2014,shi2015,noble2021,mignon2023}. Furthermore, mass exchange between the minidiscs can introduce additional periodicity in the pre-merger phase~\citep{bowen2017, westernacher-schneider2022}.

Post-merger signatures are also expected to arise from the abrupt mass loss and recoil velocity imparted to the remnant BH, caused by the burst of anisotropic GW emission at merger. These effects are nearly instantaneous, with the BH  typically losing a few per cent of its mass \citep[e.g.][]{tichy2008} and recoiling at hundreds of km~s$^{-1}$~\citep[e.g.][and references therein]{baker2008, lousto2013}, depending on the binary mass ratio, eccentricity, and BH spins. In the idealised case of a near-Keplerian, geometrically thin circumbinary disc whose mass is negligible compared with the BH remnant \citep[a good approximation for MBHBs detectable by LISA, out to radii $\sim (10^2-10^4) \times r_{\rm g}$, where $r_{\rm g }= GM_{\rm bin}/c^2$ is the gravitational radius according to the total BH mass $M_{\rm bin}$, see e.g.][]{lippai2008}, the change in the potential and the effective kick received by the gas cause spiral and/or concentric circular shock-waves to propagate outwards in the disc, as shown in analytic orbital calculations \citep[e.g.][]{lippai2008,Schnittman2008,shields2008,penoyre2018}, hydrodynamic simulations~\citep{oneill2009,corrales2010,rossi2010,megevand2009, rosotti2012}, and also in relativistic simulations of non-Keplerian discs~\citep{zanotti2010}. The resulting ``afterglows'' can emerge from weeks to months after the merger, and exhibit a characteristic time-dependent spectral energy distribution (SED).

In the present paper, we revisit the above observational signatures accompanying the merger of a pair of massive BHs.  Our motivation for this is two-fold.    

First, to discover the pre-merger signatures, the ongoing merger, detected by LISA, would need to be sufficiently well localised to allow a prompt observational campaign by large field-of-view (FOV) instruments~\citep{kocsis2008}. According to recent forecasts, sky localisation to within $\sim10$ deg$^2$ (the FOV of the Vera Rubin Observatory, see \citealt{aLSST2020}) will typically only be available one to two days before merger~\citep{mangiagli2020}. This motivates developing a better understanding of signals immediately preceding the merger. In particular, the aforementioned hydrodynamical simulations of pre-merger circumbinary disc dynamics lacked adequate spatial and temporal resolution near the merger, relied on simplified thermodynamics, and/or did not track the binary inspiral for long enough from the time of the decoupling to the merger.

Second, existing models of post-merger afterglows \citep[e.g.][]{rossi2010, corrales2010} have employed many simplifying assumptions, such as a smooth, laminar, inviscid, near-Keplerian initial disc in a Newtonian potential, and either an isothermal or an adiabatic (i.e. non-radiative) equation of state. Since these works, it has been understood that the vicinity of the binary at merger is not devoid of gas; rather there is a distorted cavity that is filled with streams of gas \citep{farris2015a,tang2018}.
At present, there are no predictions for the EM signatures arising from the mass loss and the recoil immediately following the merger from this inner region of the disturbed circumbinary disc.

In this paper, we address the above limitations by employing high-resolution hydrodynamical simulations with the fixed-mesh GPU-enabled code \texttt{Sailfish}\footnote{https://github.com/clemson-cal/sailfish} to compute the EM signatures immediately preceding and following the merger. Our work extends previous findings by incorporating physical viscosity and a more realistic treatment of thermodynamics, directly solving the energy equation, using a $\Gamma$-law equation of state for the gas, and a physically-motivated cooling prescription. Additionally, we simulate the binary inspiral for several hundred orbits, which allows us to follow the system from the fiducial decoupling epoch all the way past merger. We perform a suite of simulations varying the disc Mach number and viscosity, the binary mass, and grid resolution to test the robustness of our results. 

Our main result is a novel multi-band EM signature: we find that the fiducial $10^6~{\rm M_\odot}$ MBHB model produces strong thermal X-ray emission until one to two days prior to merger (in agreement with previous works). However, at that stage, the minidiscs are abruptly disrupted and the shocked streams disappear from the inner cavity. As a result, the X-ray luminosity drops by orders of magnitude, and the source remains X-ray dark for several days afterwards, regardless of any post-merger effects such as recoil or mass loss. At the same time, the optical and infrared emission, dominated by circumbinary gas farther out, remains nearly steady. These distinct signatures could help identify LISA counterparts without the need for localising the source on the sky more than one to two days prior to merger.

The remainder of this paper is organised as follows.  In \S~\ref{sec:Setup}, we discuss our hydrodynamics code, numerical schemes, and the initial setup for our list of runs. In \S~\ref{subsec:Fiducial} we present our main results on the gas dynamics and EM signatures before and after the merger. In \S~\ref{subsec:Params} we discuss how our results depend on varying the most important parameters in our setup. In \S~\ref{subsec:xobs} we discuss observational issues and strategies. Finally, in \S~\ref{sec:Con} we summarise our main conclusions and the implications of this work.

\section{Hydrodynamical Setup, Post-processing, and Models}
\label{sec:Setup}

\subsection{Hydrodynamical Setup}
\label{subsec:HydroSetup}

All simulations were run using the publicly-available two-dimensional GPU-enabled hydrodynamics code \texttt{Sailfish}. In this section, we give a brief summary of the technical aspects of {\tt Sailfish} \citep[for full details, see][]{westernacher-schneider2022}.

We solve the vertically-integrated Newtonian fluid equations, keeping the lowest nontrivial order in powers of $z/r$ under the conditions of a thin disc ($h/r\ll 1$) and mirror symmetry about $z=0$.\
 These equations read
\begin{eqnarray}
    \partial_t \Sigma + \nabla_j \left( \Sigma v^j \right) &=& S_{\Sigma} \label{eq:mass} \\
    \partial_t \left( \Sigma v_i \right) + \nabla_j \left( \Sigma v^j v_i + \delta^j_i \mathcal{P} \right) &=& g_i + \nabla_j \tau^j_i + S_{p, i} \label{eq:mom} \\
    \partial_t E + \nabla_j \left[ \left( E+\mathcal{P} \right) v^j \right] &=& v^jg_j + \nabla_j \left( v^i \tau^j_i \right) \nonumber\\
    &-& \dot{Q} + S_{E} \label{eq:en},
\end{eqnarray}
where $\Sigma$ is the surface density, $\mathcal{P}$ is the vertically-integrated pressure, $v^i$ is the mid-plane horizontal fluid velocity, $E=\Sigma \epsilon + (1/2)\Sigma v^2$ is the vertically-integrated energy density, $\epsilon$ is the specific internal energy density at the mid-plane of the disc, $g_i$ is the vertically-integrated gravitational force density, and $\tau^j_i = \Sigma \nu \left( \nabla_i v^j + \nabla^j v_i - (2/3)\delta^j_i \nabla_k v^k\right)$ is the viscous stress tensor (in a form that is trace-free in a 3-dimensional sense) with zero bulk viscosity, $\nu$ is the kinematic shear viscosity, $S_{\Sigma}$, $S_{p,i}$, and $S_E$ are mass, momentum, and energy sinks, and $\dot{Q}$ is the local blackbody cooling prescription, assuming hydrogen dominates the gas density \citep[see e.g][]{frank2002}, given as

\begin{equation}
    \Dot{Q} = \frac{8}{3} \frac{\sigma} {\kappa\Sigma} \left(\frac{m_p\mathcal{P}}{k_B\Sigma}\right)^4 \label{eq:cooling},
\end{equation}

\noindent
where $\sigma$ is the Stefan-Boltzmann constant, $\kappa$ = 0.4 cm$^2$ g$^{-1}$ is the opacity due to electron scattering, $m_p$ is the proton mass, and $k_B$ is the Boltzmann constant. Thermal conductivity is neglected.

\begin{comment}
    $T = (m_p/k_B)\mathcal{P}/\Sigma$.
\end{comment}

We use a torque-free sink prescription \citep{dempsey2020, dittmann2021} to model the removal of gas by each point mass. We choose the sink radius equal to $r_{\rm s}$ (where $r_{\rm s} \equiv 2GM_{\rm bh}/c^2$ is the Schwarzschild radius of a single BH). This allows us to achieve sufficient resolution near the BHs (sink diameter = 8 cells).

We initialise the disc to the conditions described in \cite{goodman2003}:

\begin{align}
    \Sigma &= \Sigma_0\left(\frac{\sqrt{r^2+r_\mathrm{soft}^2}}{a_0}\right)^{-3/5}\\
    \mathcal{P} &= \mathcal{P}_0\left(\frac{\sqrt{r^2+r_\mathrm{soft}^2}}{a_0}\right)^{-3/2}\\
    \vec{v} &= \left({\frac{GM_{\rm bin}}{\sqrt{r^2+r_\mathrm{soft}^2}}}\right)^{1/2}\hat{\phi},
\end{align}

\noindent
where $r_{\mathrm{soft}}$ is the gravitational softening length, which mimics the vertically-integrated component of the gravitational force in the plane of the disc, and is set equal to the sink radius, and $a_0$ is the initial binary separation.

Additionally, we initialise a cavity at $r=2\,a_0$ by multiplying both $\Sigma$ and $\mathcal{P}$ by a window function $f(r)$, given by:
\begin{equation}
    f(r) = 10^{-4} + (1-10^{-4})\mathrm{exp}({-2a_0/(r^2+r_\mathrm{soft}^2})^{1/2})^{30}.
\end{equation}
We use a $\Gamma$--law equation of state $\mathcal{P}=\Sigma\epsilon(\Gamma-1)$, where $\Gamma=5/3$. While the gas is more likely to be radiation-dominated ($\Gamma=4/3$), our choice allows us to compare with previous results most closely related to our study \citep[][which all used $\Gamma=5/3$]{corrales2010,farris2015a,tang2018,westernacher-schneider2022}, and to avoid the viscous and thermal instabilities associated with such $\Gamma=4/3$ models \citep[][respectively]{lightman+1974, shakura+1976}. We use a Shakura-Sunyaev viscosity prescription $\nu=\alpha c_{\rm s}h$ \citep{shakura1973} with a fiducial $\alpha=0.1$. The sound speed is determined by $c_{\rm s}^2=\Gamma\mathcal{P}/\Sigma$. The disc's half thickness, $h$, is determined by $h=\sqrt{\mathcal{P}/\Sigma}/\tilde{\Omega}$ with $\tilde{\Omega}=\sqrt{GM_1/r_1^3+GM_2/r_2^3}$, where $r_1$ and $r_2$ are the  distances from the particle and the respective black holes. The disc mass is small compared to the mass of the binary, so self-gravity can be ignored.

Although hydrodynamic radiation pressure is typically important in the innermost regions of MBH discs, including it in simulations is technically challenging. Thus, we omit radiation pressure, and instead initialise our disc with a characteristic disc aspect ratio $h/r$ that is similar to that expected when radiation pressure is accounted for. In practice, the characteristic disc aspect ratio is adjusted via choices of $\Sigma_0$ and $\mathcal{P}_0$, and when interpreted literally, can imply an extremely super-Eddington accretion rate~\citep{dorazio2013}. The effective temperature of the disc, $T_{\mathrm{eff}}$, is related to mid-plane temperature $T$ by
\begin{equation}
    T_{\mathrm{eff}}^{4} = \frac{4}{3}\frac{T^4}{\kappa\Sigma}
    \label{eq:teff},
\end{equation}
In \S~\ref{subsec:Post}, we describe how in post-processing we scale the effective temperature down into the range expected when radiation pressure is accounted for. Although the disc aspect ratio develops self-consistently from a balance of heating and cooling in the simulation runs, we find its initialised characteristic value is representative of the fully-developed system. We relate the accretion rate to the surface density and kinematic shear viscosity via $\dot{M}=3\pi\Sigma\nu$ \citep{frank2002}.

The combined mass of the binary was chosen to be $M_{\rm bin}=10^6$~\(\rm M_\odot\), roughly matching the value where LISA is most sensitive \citep{amaro-seoane2017}. Simulations suggest that gas accretion drives binaries towards equal mass, \citep[see e.g.][]{farris2014,duffell2020} so we simulate binaries with mass ratios $q\equiv m_2/m_1 = 1$. An initial separation of $a_0=50\,r_{\rm g}$ (where $r_{\rm g} \equiv GM_{\rm bin}/c^2$ is the gravitational radius for the total binary mass) was chosen to be wide enough for the binary to traverse the fiducial decoupling radius.
More specifically, the initial viscous time, $t_{\rm \nu} = 2/3\ r^2 / \nu$, in our fiducial run at $r=a_0$ is 16.54 days, which is shorter than the inspiral time of 27.84 days. The binary orbits are circular and the BHs are assumed to have zero spin.

The gravitational field of each individual BH is modelled by a Plummer potential,
\begin{equation}
    \Phi_n = -\frac{GM_n}{\sqrt{r_n^2+r_{\mathrm{soft}}^2}},
\end{equation}
where $M_n$ is the mass of the $n$th BH, $r_n$ is the distance from a field point to the $n$th black hole, and $r_{\rm soft}$ is the gravitational softening length scale introduced above. A pseudo-Newtonian potential, such as Paczy\'nski-Wiita \citep[][]{PW1980}, would likely benefit our study. However, when attempting to employ this potential, we noticed post-merger numerical effects in which the sinks did 
not appear to be removing material properly. As such, we instead opted to use a Plummer potential. We intend to resolve these numerical issues in future work, and perhaps explore more advanced pseudo-Newtonian potentials, in addition to Paczy\'nski-Wiita.

The realistic state of the disc, including the eccentric cavity and lump in the cavity wall, takes time to develop from our idealised axisymmetric initial conditions. Thus, before initiating inspiral, we allow our system to evolve for 820 orbits on a fixed circular orbit, corresponding to $\sim$6.3 viscous times at $r=a_0$. By comparison, the time to merger is roughly 25\% of this duration (approximately 220 initial orbital times).

The inspiral is implemented with the quadrupole approximation
\citep{Peters1964}, i.e. the binary separation follows
\begin{equation}
    a(t) = a_0(1-t/\tau)^{1/4}
    \label{eq:peters},
\end{equation}
\noindent
where $\tau = a_0^4/4\beta$ is the total GW inspiral time and $\beta\equiv(64/5)(G^3/c^5)m_1m_2(m_1+m_2)$. For the binary separation of $50~r_{\rm g}$, this implies $\tau\sim28$ days $\sim$ 220 initial orbits $\sim$ 350 total orbits. This initial separation ensures that we follow the binary through the decoupling limit. Once the BHs are within two Schwarzschild radii of each other, the BHs are instantly merged, with a new position at the origin and a sink radius of 2~$r_{\rm s}$, double the radius of either initial BH.

Recoil is expected to occur nearly instantaneously at merger. Thus, in our kicked simulations, immediately after the merger an in-plane velocity of 530 km s$^{-1}$ is applied to our BH. This is a typical value, and also what was chosen in \citet{corrales2010}, allowing our results to be compared to that work. Similarly, as mass loss is also expected to occur approximately instantaneously at merger, we treat it as such, by reducing the total mass of the resulting BH by $3\%$ for our mass-loss simulations, in line with expected values. We examine kicks in four different directions with respect to the orientation of the cavity, as described below, as well as a reference model without any kick. All of these setups are examined with and without mass loss. These variations, along with the additional model runs discussed in \S~\ref{subsec:Models}, are listed in Table~\ref{tab:Vars}.
We use a uniform Cartesian grid with a square domain of side length $40~a_0$ and $4000^2$ cells, giving a resolution of $\Delta x=\Delta y=0.01\,a_0$. We use a buffer source term at the outer boundary off the grid, the purpose of which is to drive the solution to the initial conditions for the disc, preventing the outer boundary from propagating artefacts inward. Additional tests include: changing the sink prescription inside the sink to immediately set fluid velocities to zero and the pressure and density to their respective floor values; altering the density and pressure floor values; and running with different PLM (piece-wise linear method) values \footnote{This controls the aggressiveness of slope-limiting in the hydrodynamic solver.}.
All of these tests reveal negligible effects on our scientific conclusions. We also perform a resolution test in our fiducial model to ensure convergence of our results (Appendix \ref{app-a}).

\subsection{Post-processing}
\label{subsec:Post}

In post-processing, the effective temperature is obtained from Eqs.~\eqref{eq:cooling} and~\eqref{eq:teff}, leading to

\begin{equation}
    \Dot{Q}=2\sigma T_{\mathrm{eff}}^{4},
\end{equation}
\noindent
where the factor of 2 comes from the fact that the disc cools through two faces (top and bottom). Since the effective temperature is related to the accretion rate via
\begin{eqnarray}
    T_{\rm eff}^4 \propto \dot{M},
\end{eqnarray}
the artificially high accretion rates required to achieve realistic characteristic disc aspect ratios in the gas-dominated models result in artificially high effective temperatures, which affect the EM luminosity and spectrum. Therefore, we correct for this in post-processing by uniformly re-scaling the effective temperature back down to our target system via the map $T_{\rm eff}^4 \rightarrow T^4_{\rm eff} / M_{\rm boost}$, where $M_{\rm boost}$ is the dimensionless ratio of the accretion rate of the gas pressure model to that of the ``target'' model where radiation pressure is accounted for.

We assume black-body emission from each cell of our domain, which allows us to compute the luminosity in different bands. We neglect Doppler effects, so our light curves are valid for observers who are viewing the disc face-on. We discuss how our results are modified when the light curves are modulated by relativistic Doppler boosting in Appendix \ref{app-b}. We calculate the luminosity of an area element $dA$ in the frequency band between $\nu_1$ and $\nu_2$ via
\begin{equation}
    dL=\pi dA \int_{\nu_1}^{\nu_2}\frac{2h\nu^3/c^2}{ \mathrm{exp}\left(\frac{h\nu}{kT_\mathrm{eff}}\right)-1 }d\nu,
\end{equation}
To obtain the total luminosity of the disc, we sum over the spatial domain, excluding the outer buffer region. We examine the bolometric luminosity, as well as the luminosities in three fixed bands $E_{\rm opt}:1.8-3.1~\rm eV$, $E_{\rm UV}:3.1-124.0~\rm eV$, $E_{\rm X-ray}:124.0~\rm eV -124.0~\rm keV$.
\noindent

\subsection{Suite of Models}
\label{subsec:Models}

We also explore variants of our model with an increased viscosity of $\alpha=0.3$, an increased total binary mass of $M_{\rm bin}=10^8$~\(\rm M_\odot\), and an increased Mach number of $\mathcal{M} = 30$. In each of these models we again examine the cases with no kick, kicks in four directions, and with and without mass loss in every case. Our full suite of models is shown in Table~\ref{tab:Vars}.

\begin{table*}
\caption{Our suite of simulations. The bold denotes which parameter has been modified from the fiducial model in the first row. All models listed are run with no kick, kicks in four different directions with respect to the orientation of the cavity, and with and without mass loss
in every case, *except Res\_8k, which was only run in the most luminous kick direction (described in \S~\ref{subsubsec:PostEffects}) . When applied, the recoil velocity is 530 km s$^{-1}$ and the mass loss is $3\%$ of the binary's total initial mass for all simulations.}
\centering
\begin{tabular}{ p{0.12\linewidth} | P{0.2\linewidth} | P{0.12\linewidth} P{0.12\linewidth} P{0.12\linewidth} P{0.12\linewidth}}
 \hline \hline
 Model &Description &$\alpha$ &Mass $($M$_\odot)$ &Mach &Res $(\Delta x)$\\
 \hline
 Fiducial &Fiducial Model &0.1 &$10^6$ &10 &0.01\\
 Alpha0.3 &Higher viscosity &\boldmath{$0.3$} &$10^6$ &10 &0.01\\
 Mass$10^8$ &Higher BH mass &0.1 &\boldmath{$10^8$} &10 &0.01\\
 Mach30 &Higher Mach number &0.1 &$10^6$ &\boldmath{$30$} &0.01\\
 Res\_1k &Lowest resolution  &0.1 &$10^6$ &10 &\boldmath{$0.04$}\\
 Res\_2k &Lower resolution &0.1 &$10^6$ &10 &\boldmath{$0.02$}\\
 Res\_8k* &Highest resolution &0.1 &$10^6$ &10 &\boldmath{$0.005$}\\
 \hline
 \hline
\end{tabular}
\label{tab:Vars}
\end{table*}

\section{Results}
\label{sec:Results}

In this section, we describe the results of our simulation library, first our fiducial model, and then how our main results depend on model parameters. 

\subsection{Fiducial model}
\label{subsec:Fiducial}
We present in this sub-section the main results from the fiducial merger. We summarise the evolution up to the merger and how this affects the EM emission, then highlight post-merger effects and comment on the nature of the periodicities we find in EM light curves. 

\subsubsection{Binary evolution up to merger}

\begin{figure*}
    \centering
    \includegraphics[width=0.9\textwidth]{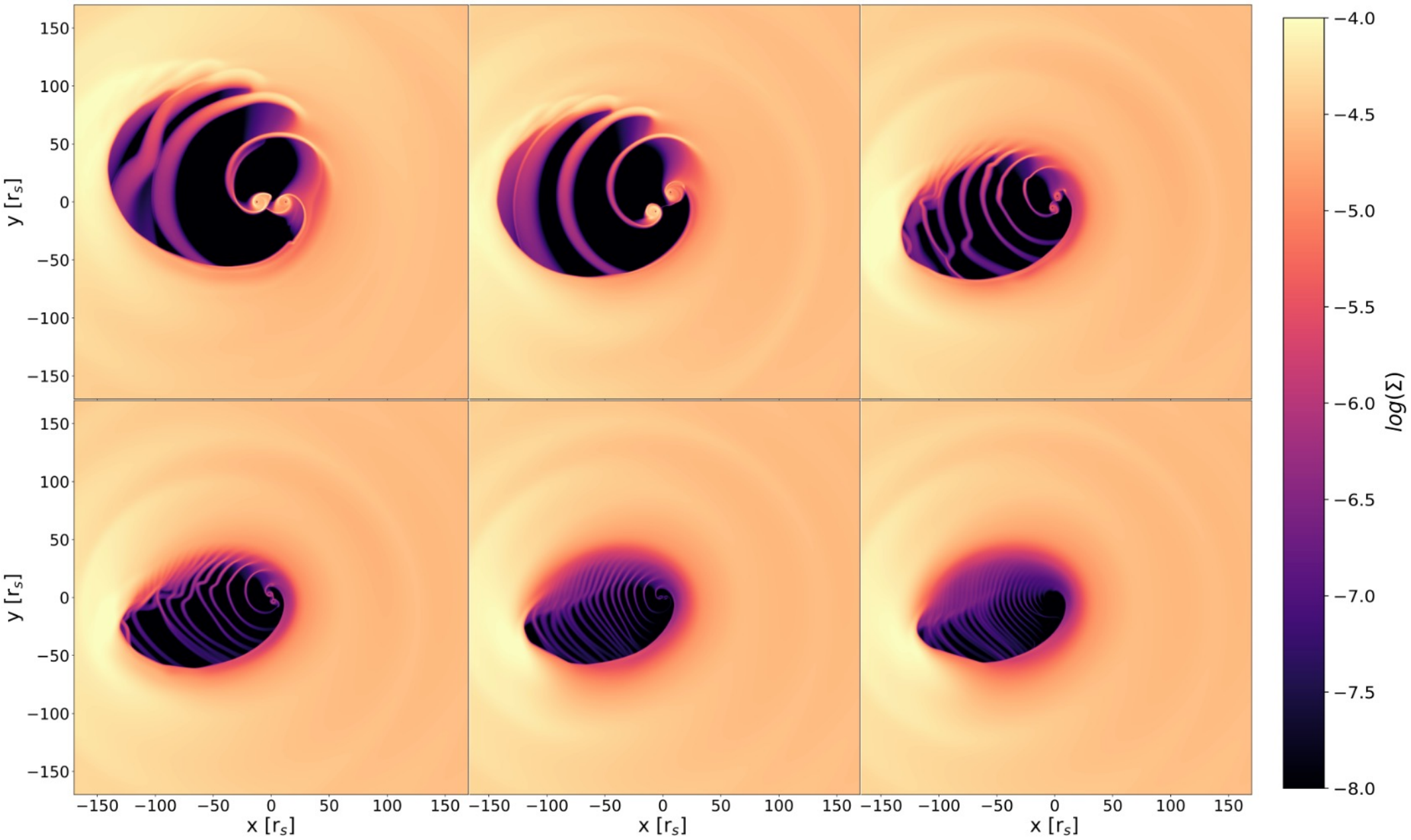}
    \caption{Snapshots of the logarithmic surface density from the beginning of inspiral to merger, shown (reading left to right, top to bottom) at 27.8 days, 13.9 days, 1 day, 5 hours and 1 hour before merger, and at merger.}
    \label{fig:fid_inspiral_to_merger}
\end{figure*}

After running the simulation for about six viscous times, the CBD is in a settled state. Figure~\ref{fig:fid_inspiral_to_merger} shows the logarithmic surface density for the $\sim$inner third of our domain. The top left panel, which shows the system just before inspiral is initiated, shows a clear eccentric cavity, represented by the under-dense region which surrounds the binary. As the black holes orbit they pick up mass from the inner edge of the cavity, creating streams within the cavity. Parts of these streams are accreted onto two smaller accretion discs around each black hole, the minidiscs, while other parts are expelled towards the opposite side of the cavity. The minidiscs are clearly visible as overdensities around each black hole.   All of these features have been found in previous simulations of accreting equal-mass binaries on fixed circular orbits.  

The other five panels in Figure~\ref{fig:fid_inspiral_to_merger} show the later stages of the binary inspiral, ending with the merger in the bottom right panel. As binaries orbit, the minidisc tidal truncation radius tracks, on average, $\sim1/3$ of the instantaneous separation \citep{paczynski1977}. As the binary shrinks, the minidisc tidal truncation radius also shrinks, generally leading to a decrease in their total mass. The frequency of stream creation naturally increases with the binary frequency, producing an increasing number of ejected streams visible in the cavity at later times. At merger, the minidiscs are disrupted and disappear.

\begin{figure}
    \centering
    \includegraphics[width=0.45\textwidth]{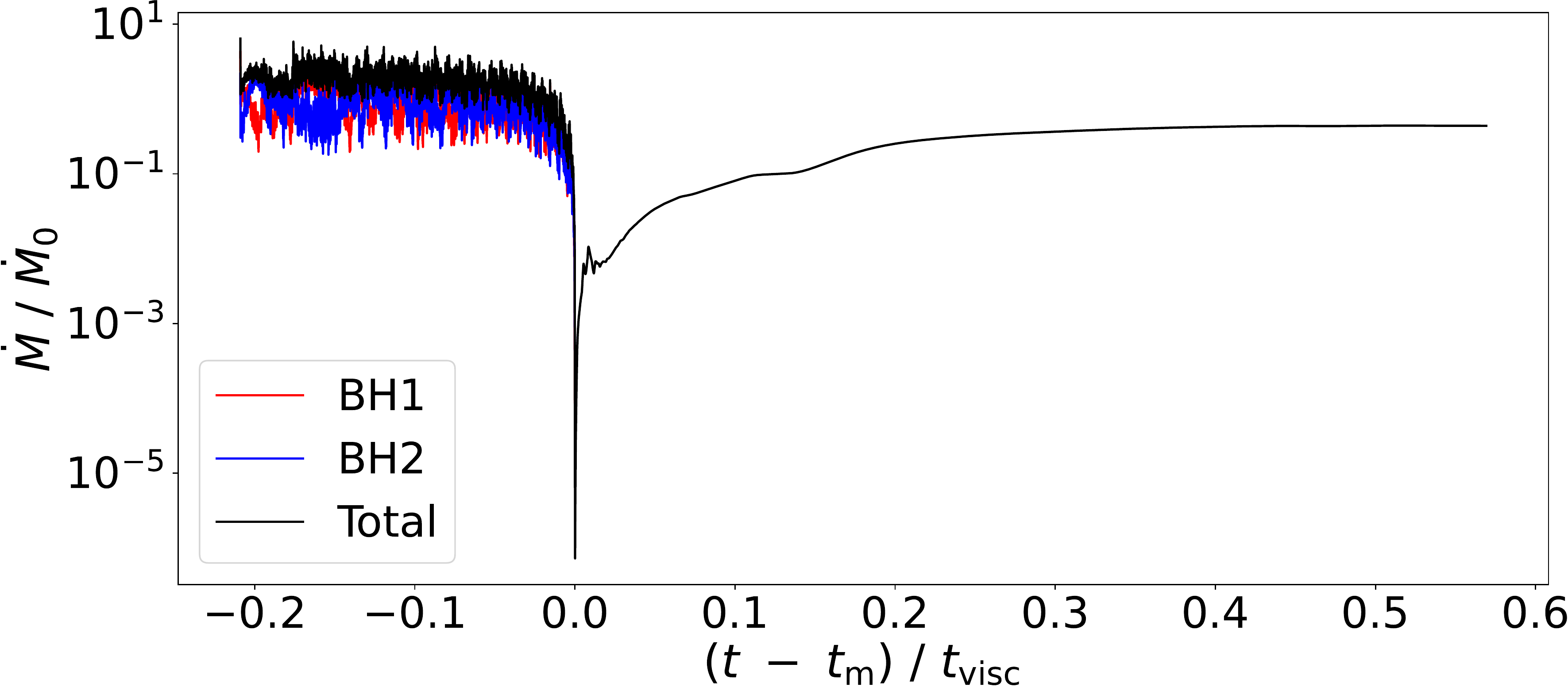}
    \caption{Black hole accretion rates before and after the time of merger, $t_m$, for the case with no post-merger recoil or mass loss. Accretion rates are normalised by the steady Shakura–Sunyaev value around a single BH.}
    \label{fig:fid_long_term_accretion}
\end{figure}

The mass accretion rate of both binary components is shown as a function of time in Figure~\ref{fig:fid_long_term_accretion}. As our BH would otherwise move out of the simulation domain in this time period, the model shown has no kick (nor mass loss).
A significant drop in the accretion rate occurs at the time of merger. This indicates that as the minidiscs surrounding the BHs disperse, the accretion also abates, at least temporarily. Accretion begins to restore soon after merger, and appears to settle over nominal viscous time scales.

\begin{figure}
    \centering
    \includegraphics[width=0.45\textwidth]{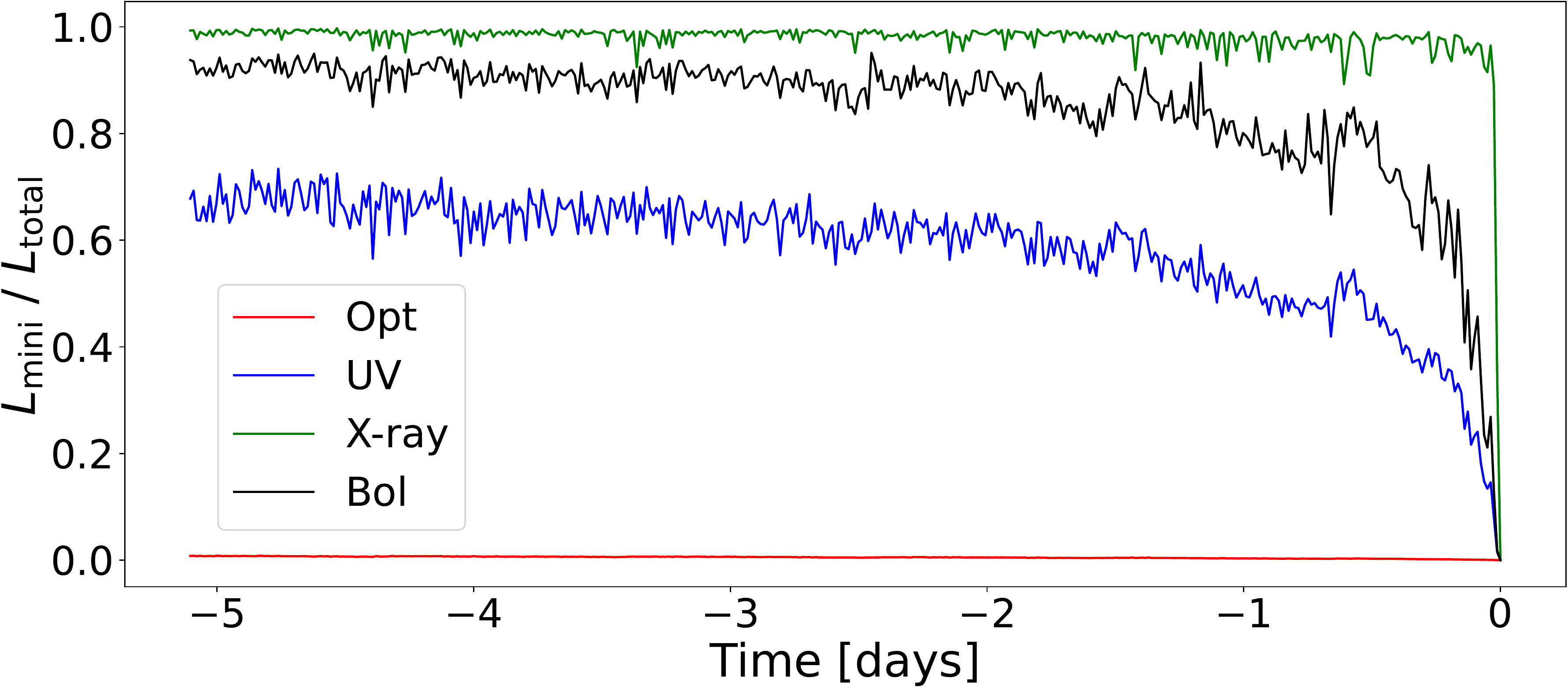}
     \caption{The fraction of the total luminosity contributed by the minidiscs in different bands, as labelled. The minidiscs produce nearly all of the X-rays but almost none of the optical emission.}
    \label{fig:fid_band_ratios_5days}
\end{figure}

\begin{figure*}
    \centering
    \includegraphics[width=0.9\textwidth]{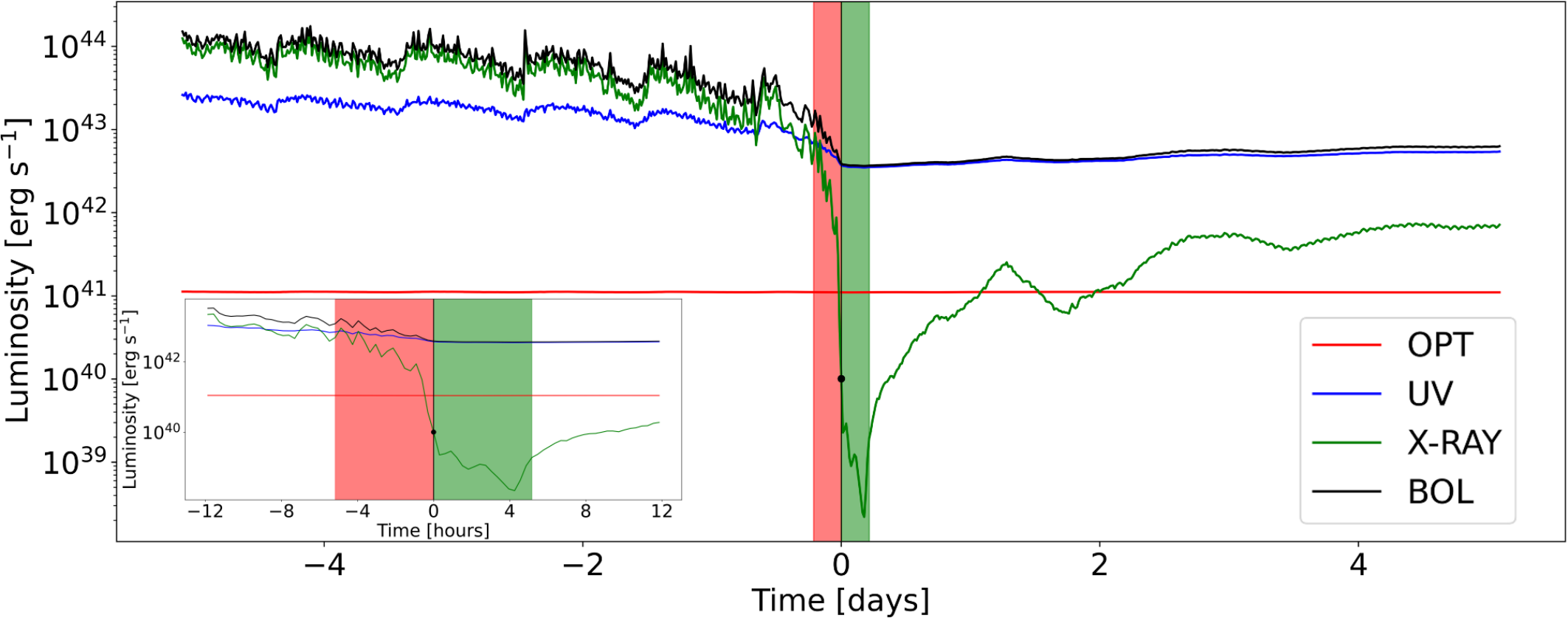}
\caption{Optical, UV, X-ray, and bolometric light curves from five days before merger to five days after merger. The black vertical line is centred at the merger ($t=0$), with a black dot marking the X-ray luminosity at the moment of the merger. The red/green zones indicate five hours before/after the merger. The inset shows a zoom-in version around the merger.  This run includes a recoil kick (pointed downward in the bottom right panel of Fig.~\ref{fig:fid_inspiral_to_merger}), contributing to the post-merger recovery of the luminosity, but does not include mass loss, which inhibits recovery.}
    \label{fig:fid_bands_5days}
\end{figure*}

Since mass accretion rates are not directly observable, it is interesting to investigate in which EM bands this minidisc depletion could leave an imprint. We first investigated in which EM frequency band the minidiscs are brightest. Figure~\ref{fig:fid_band_ratios_5days} shows the fraction of optical, UV, X-ray, and bolometric luminosities that originate from the minidiscs. Luminosity is considered to have originated from the minidiscs if it is emitted from a circular region centred on the binary barycentre, whose radius is chosen to lie just outside the outer-most part of either minidisc (calculated from a combination of the instantaneous binary separation and truncation radius). The bulk of the bolometric emission (black curve) comes from the minidiscs. Practically all of the X-ray emission (green curve) comes from the minidiscs, whereas practically all of the optical emission (red curve) comes from the CBD. At the moment of merger, a clear drop in the minidiscs' fraction of X-ray emission is evident. Since the minidiscs survive up until hours before the merger, and they dominate the X-ray emission, their destruction leads to a corresponding drop in X-ray emission. This can be seen in Figure~\ref{fig:fid_bands_5days}, which shows light curves before and after merger for a case with a post-merger kick. The inset shows a zoomed-in view of the 12 hours before and after merger. The red and green shaded regions indicate 5-hour windows before and after the merger. A drop of five orders of magnitude is seen in the X-ray luminosity during this period. In the post-merger phase, partial re-brightening of X-rays can arise from shocks formed in the CBD due to black hole recoil. We consider such post-merger effects in detail next.

\subsubsection{Post merger effects}
\label{subsubsec:PostEffects}

\begin{figure*}
    \centering
    \includegraphics[width=0.9\textwidth]{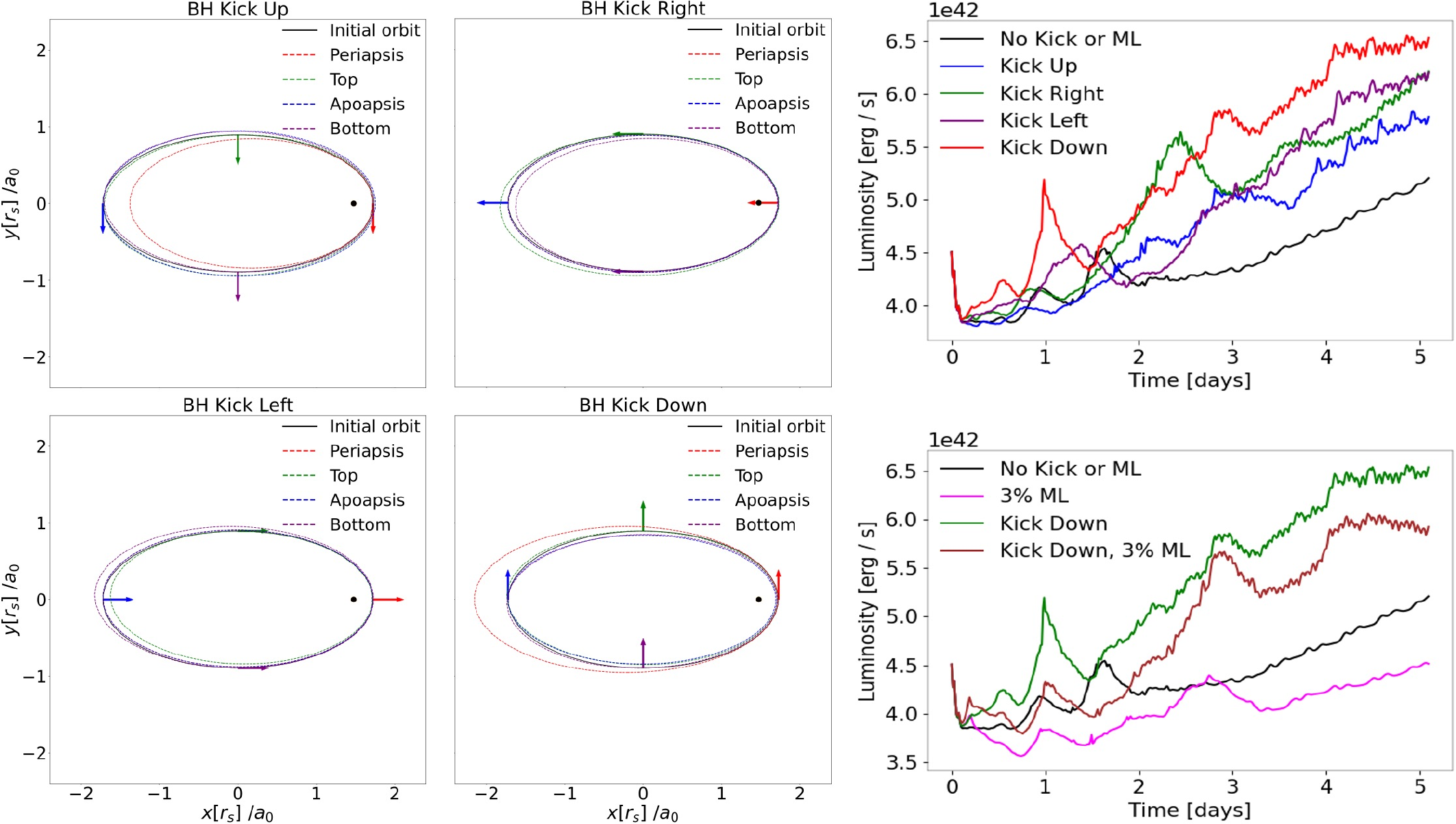} 
    \caption{The left four panels correspond to four different kick directions for the BH at merger (Up, Right, Left, Down). In each panel, different colours show the perturbed post-merger orbital paths of test particles which suffered a recoil kick at four different positions along their initial orbit (at Periapsis, Top, Apoapsis, and Bottom). The top right panel shows the simulated post-merger bolometric light curves in our fiducial model with the same four different kick directions as well as no kick (or mass loss) for comparison. This panel demonstrates that the highest (red) and lowest (blue) post-merger luminosities are produced by kicks in the Down and Up directions. As the left panel shows, these are the directions which re-direct test particles near their periapsis towards and away from the cavity wall, respectively. The bottom right panel shows the bolometric light curves for no kick or mass loss, 3\% mass loss, kick in the downwards (most luminous) direction, and both a kick in the downwards direction and a 3\% mass loss. This panel demonstrates that mass loss tends to reduce the post-merger luminosity, both in the absence and the presence of a kick.}
 \label{fig:fid_orb_mech_and_kick_lums}
\end{figure*}

We investigate post-merger effects by imparting mass loss as well as kicks in various directions. To inform our expectations for the effect of kicks, we first consider a toy model, following a test particle on a Keplerian elliptical orbit, and kick it at different points along its orbit, corresponding to periapse, apoapse, and two points in between. At each point, we examine kicks in four directions, to obtain a total of 16 kicked orbits. These orbits do not capture hydrodynamical effects, but serve to guide expectations of what happens in the simulations. The left four panels of Figure~\ref{fig:fid_orb_mech_and_kick_lums} show the corresponding trajectories. Note that Galilean invariance implies, for example, that a BH kick upward is equivalent to kicking the test particle downward (upper left panel). Particle trajectories are counter-clockwise, with kicked trajectories shown as dashed lines with varied colours. 

Taking the BH kick Up direction as an example, we see that the particles kicked at the Top, Apoapsis, and Bottom points have only slight perturbations to their trajectories. However, the particle kicked at the Periapsis point, which receives a relative kick opposing its velocity, has a more substantial change to its orbit. In particular, the orbit becomes more circular, with a significant reduction in the apoapsis. We might expect that a fluid element with this fate would be diverted away from the cavity wall, towards the inner region of the low-density cavity. Thus, we might also expect this to be the least luminous post-merger kick direction, as there are decreased chances of interactions between the gas streams. Conversely, looking at the lower right panel, BH Kick Down, the particle at Periapsis now receives a boost parallel to its velocity, which 
significantly increases its apoapsis.  We expect that in the simulation, this corresponds to fluid elements travelling deeper into the opposing wall of the circumbinary cavity. These additional collisions might in turn be expected to cause a greater overall luminosity. Finally, the luminosity resulting from the Left and Right kicks might be expected to be in between the Up and Down cases, and, because of their overall symmetry, may have similar values. 

These expectations are borne out by our hydrodynamic simulations. The upper right panel shows light curves for the same four kick directions (determined with the respect to the cavity orientation). Mass loss leads to overall lower post-merger luminosity, as seen in the lower right panel, in which we show the no-kick and kick-down results with and without mass loss.

We note that Figure~\ref{fig:fid_bands_5days} shows the case where the BH kick direction opposes the orbit of the gas at periapsis (i.e. fluid elements near periapsis receiving boosts to their velocities in the frame of the kicked BH)
and without mass loss. According to the above, this case exhibits maximal re-brightening in the post-merger phase. Yet, even in this optimal scenario, the drop in the X-ray band is evident.

\begin{figure}
    \centering
    \includegraphics[width=0.45\textwidth]{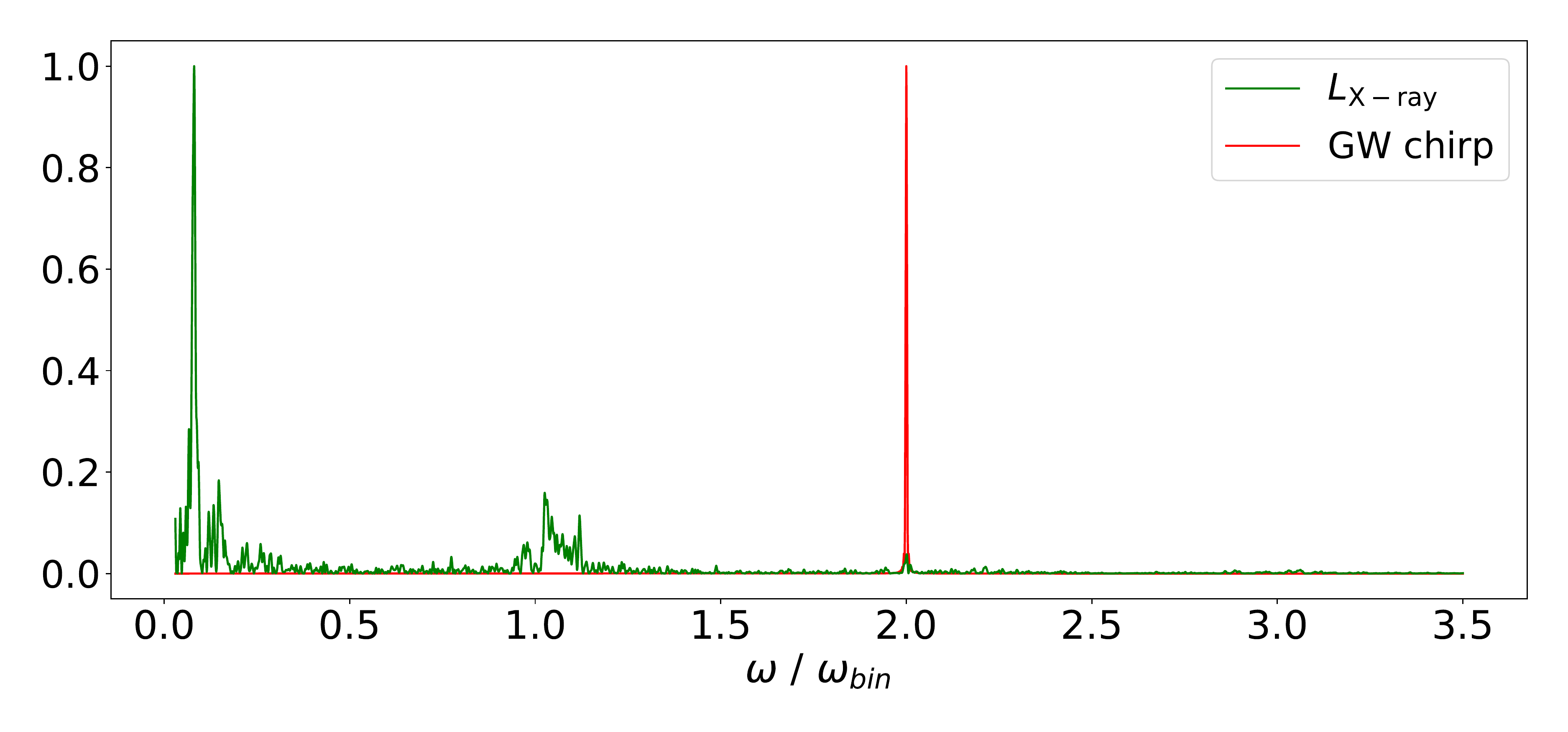}
  \caption{Lomb-Scargle periodograms, with maximum peaks normalised to unity, for the X-ray light curve (green) and GW chirp (red) during the 15 days before merger. We adopt a scaled frequency for the light curves, i.e. frequency is measured in units of the instantaneous binary orbital frequency.}
    \label{fig:pgram_15to0}
\end{figure}

\begin{figure*}
    \centering
    \includegraphics[width=0.9\textwidth]{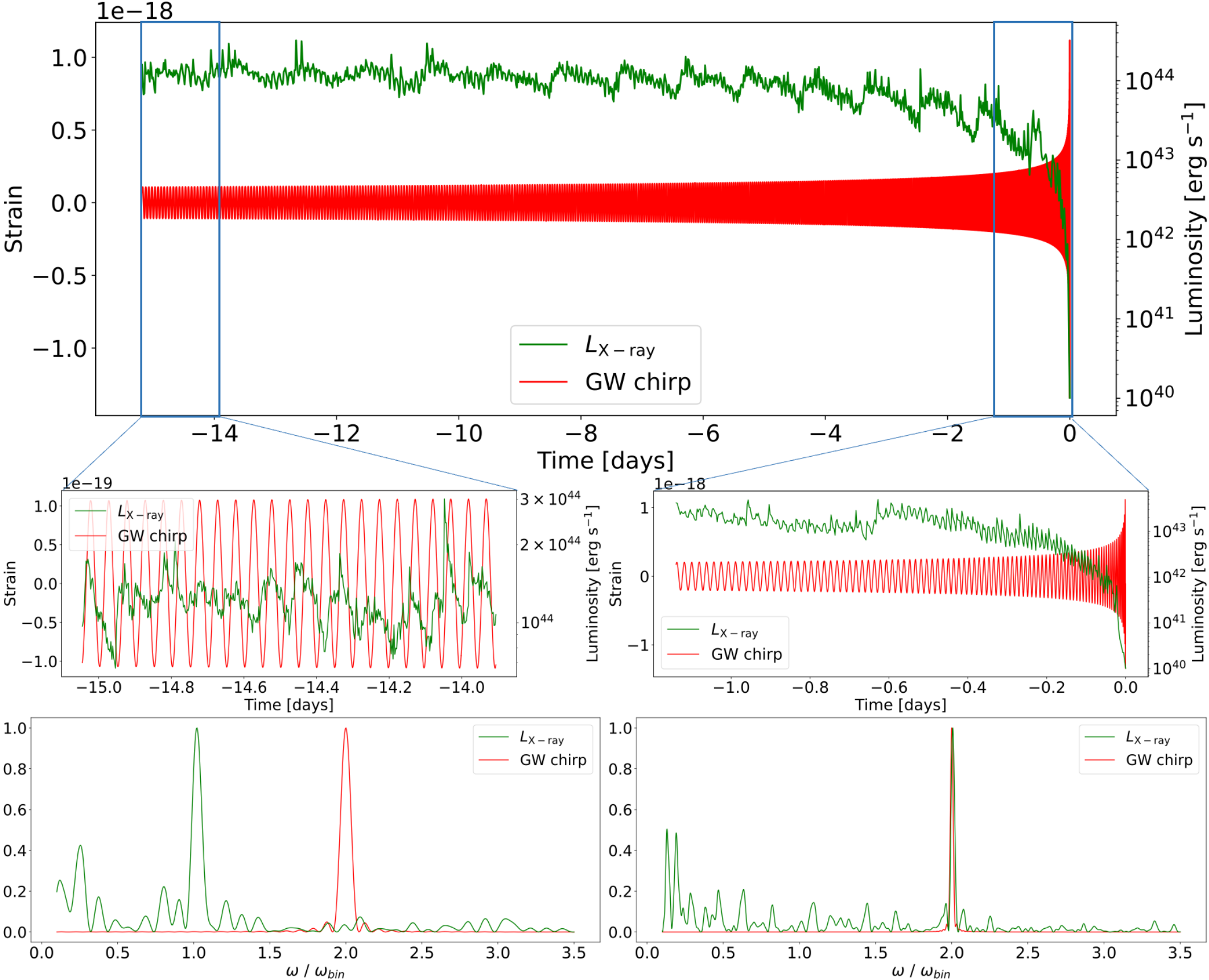}
    \caption{X-ray luminosity (green) and GW strain (red) 15 to 0 days  (top), 15 to 14 days (middle left)  and 1 to 0 days (middle right) before merger. Lomb-Scargle periodograms are shown on the lower left and right panels, with maximum peaks normalised to unity, for the X-ray light curve (green) and the GW chirp (red) during the same two time windows as the row above. Frequency is displayed in units of the instantaneous binary orbital frequency.}
    \label{fig:combined_strain_lum_pgram}
\end{figure*}

\subsubsection{Periodicities}

Figure~\ref{fig:pgram_15to0} shows the Lomb-Scargle periodogram of the X-ray light curve (green) and the GW chirp (red) for the last 15 days before merger. The GW chirp is computed from the \citealt{Peters1964} quadrupole formula for a circular equal-mass binary at $z=1$. Time is re-scaled by the instantaneous binary orbital period, resulting in a periodogram where the frequency is re-scaled by the the instantaneous binary orbital frequency. This is done to assess the periodicities of our signal in relation to the binary’s shrinking orbital period. We note three distinct peaks in the X-ray periodogram: a peak near $\omega / \omega_{\rm bin} \approx 0.1$, a peak at or slightly above the binary frequency $\omega / \omega_{\rm bin} = 1$, and a small peak in line with the GW chirp at twice the binary orbital frequency $\omega / \omega_{\rm bin} = 2$. We can investigate these frequencies by studying different segments of the light curve.

Figure~\ref{fig:combined_strain_lum_pgram} shows the X-ray light curve (green) and the GW chirp (red) for the last 15 days of inspiral in the top panel,  zoomed-in views of the first and last day of this interval in the middle two panels, and Lomb-Scargle periodograms during these two 1-day windows (with frequency measured in units of the instantaneous binary orbital frequency and the low-frequency spike omitted) in the bottom two panels. 
In the top ``panoramic'' panel, we see two distinct periodicities: a longer one on the order of a day, and a shorter one on the order of hours to minutes, both of which shorten over time. 

The longer period is produced by a well-known overdensity in the cavity wall, called a lump, which propagates a wave pattern along the cavity wall in the prograde direction. The lump forms from streams ejected from the binary, building an asymmetric gas distribution on the opposing cavity wall. The associated wave pattern propagating along the cavity wall modulates the distance between the binary and the nearest side of the cavity wall, thereby modulating the rate of CBD feeding to the binary. When CBD feeding to the binary is in a trough or a crest, the luminosity is minimal or maximal, respectively. This periodic process accounts for the low-frequency spike at $\omega/\omega_{\rm bin}= 0.1$ in the initial 15-day periodogram.

The faster X-ray periodicities in Figure~\ref{fig:combined_strain_lum_pgram} reflect mass trading activity between minidiscs, which cause EM flares. Examining the middle left panel we see that the X-ray light curve has a dominant frequency, which is evidently a near-orbital frequency $\omega/\omega_{\rm bin}\approx 1$ in the corresponding periodogram. In the middle right panel, the dominant frequency is even faster, and becomes more rapid towards merger. The corresponding periodogram in the lower right panel implicates $\omega/\omega_{\rm bin} = 2$, agreeing with the GW chirp frequency.

Visual inspection of the gas evolution in the minidiscs suggests that pulses of mass-trading between minidiscs correspond to both the near-orbital and twice-per-orbit flares. At large separation, near-orbital cadence was observed with $\Gamma$-law gas models in recent work \citep[][]{westernacher-schneider2022, westernacher-schneider2023}, and was evident in accretion rates even earlier \citep[][]{farris2015}. On the other hand, two flares per orbit in the late stages of the inspiral was reported previously with 2D simulations \citep[e.g.][]{tang2018}, and a minidisc mass-exchange mechanism with twice-per-orbit cadence was found in 3D simulations of inspiral in relativistic potentials \citep[][]{bowen2017}. Given these past findings, it is natural to expect a transition between these two cadences during inspiral. This transition is also observed in some of our sensitivity tests, but we leave a more detailed investigation of this phenomenon to future work.

\subsection{Parameter dependence}
\label{subsec:Params}

\begin{figure}
    \centering
    \includegraphics[width=0.45\textwidth]{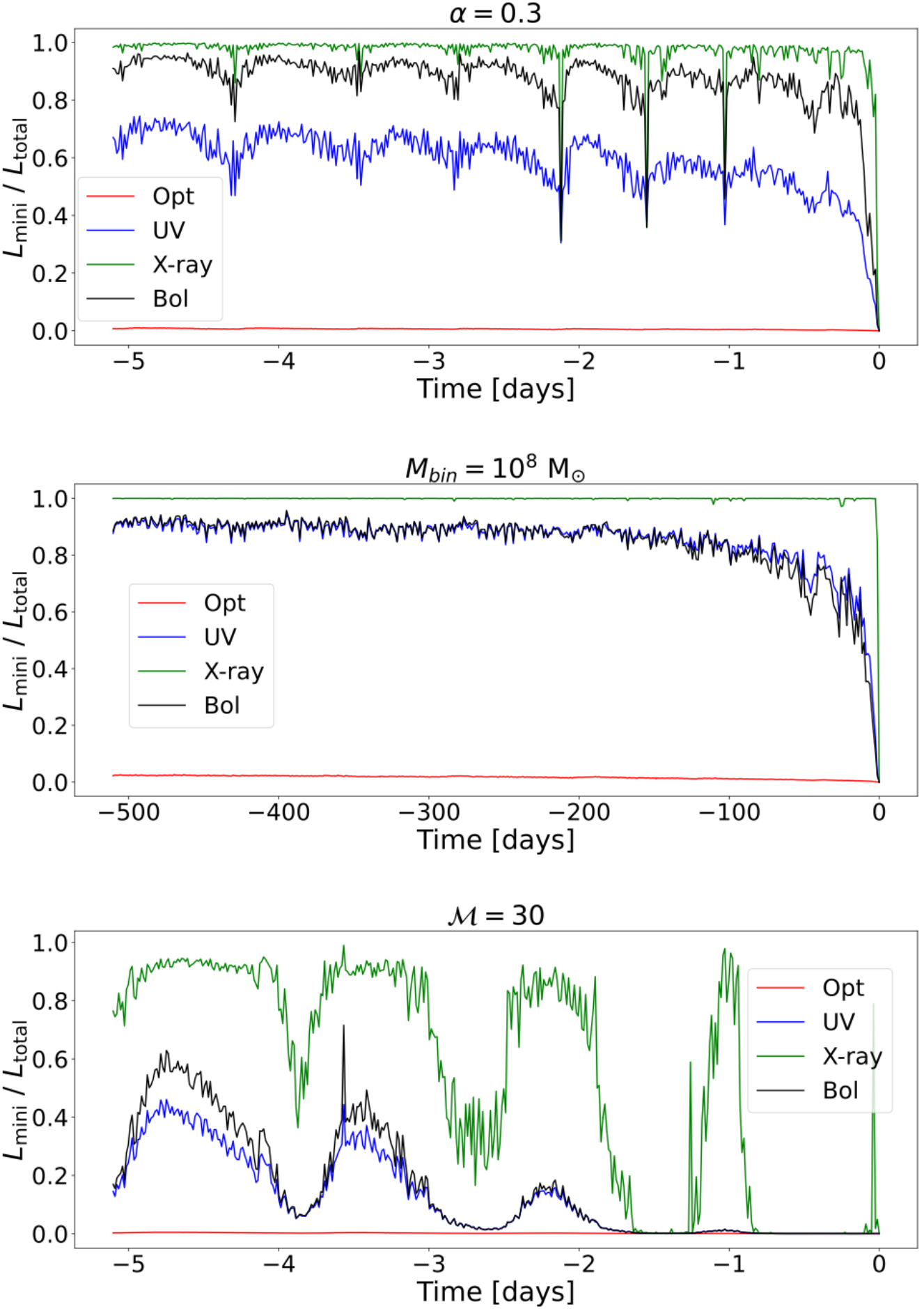}
    \caption{The fraction of each EM band originating from the minidiscs for the 
    models with a higher viscosity ($\alpha=0.3$, top), higher BH mass ($M_{\rm bin}=10^8~\rm M_\odot$,  middle) and higher Mach number ($\mathcal{M} = 30$, bottom). }
    \label{fig:bands_ratio_variations_combined}
\end{figure}

\begin{figure}
    \centering
    \includegraphics[width=0.45\textwidth]{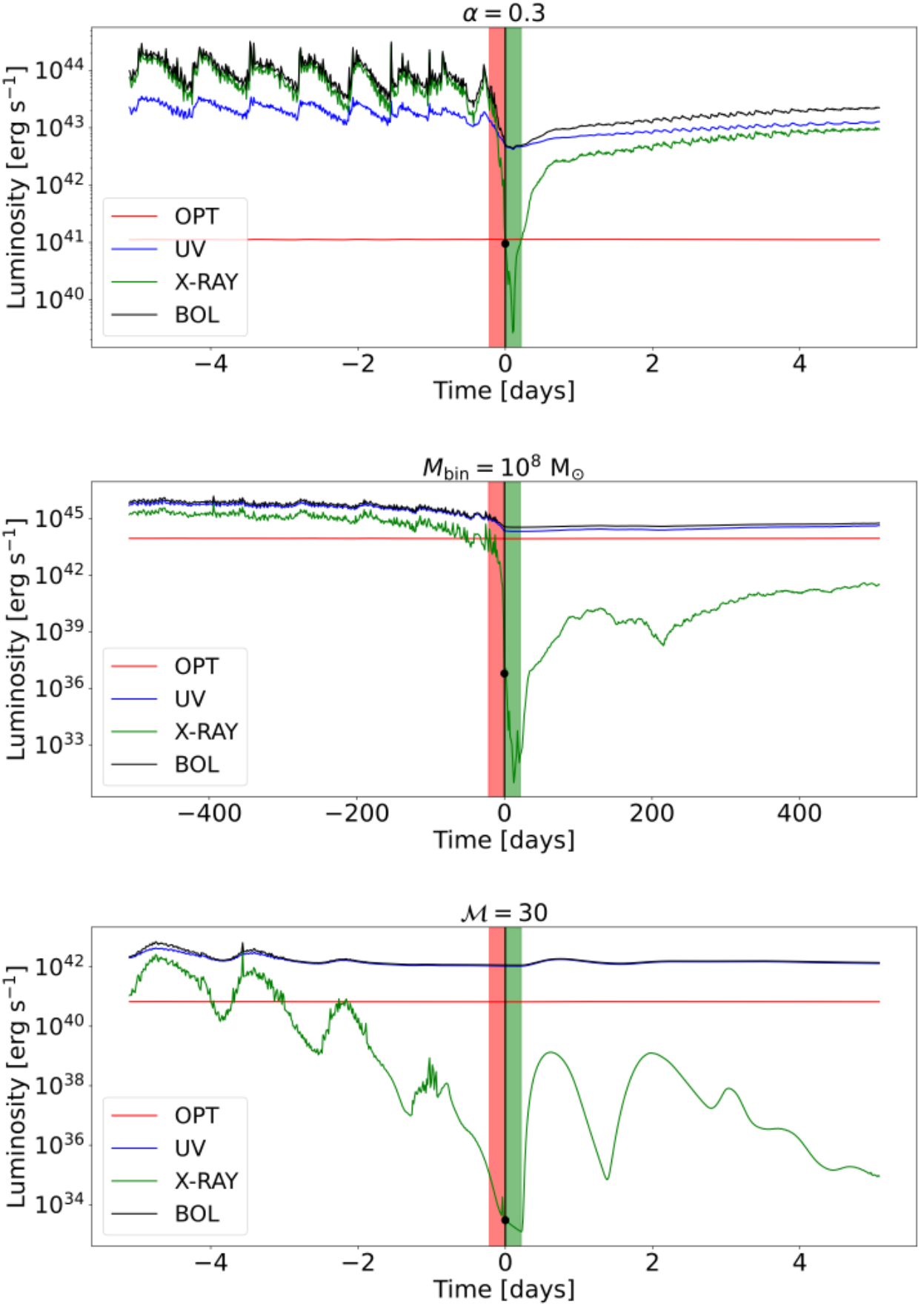}
    \caption{Luminosity in various bands for the 
    models with a higher viscosity ($\alpha=0.3$, top), higher BH mass ($M_{\rm bin}=10^8~\rm M_\odot$,  middle) and higher Mach number ($\mathcal{M} = 30$, bottom). Top and bottom panels show 5 days before and after the merger, and the red/green zones indicate 5 hours before/after the merger. Due to the different time scales for the high-mass model, the middle panel shows 500 days before and after the merger, and the red/green zones indicate 500 hours before/after the merger. In all panels, a black vertical line is centred at $t=0$, with a black dot indicating the X-ray luminosity at the moment of merger.}
    \label{fig:bands_variations_combined}
\end{figure}

\begin{figure}
    \centering
    \includegraphics[width=0.45\textwidth]{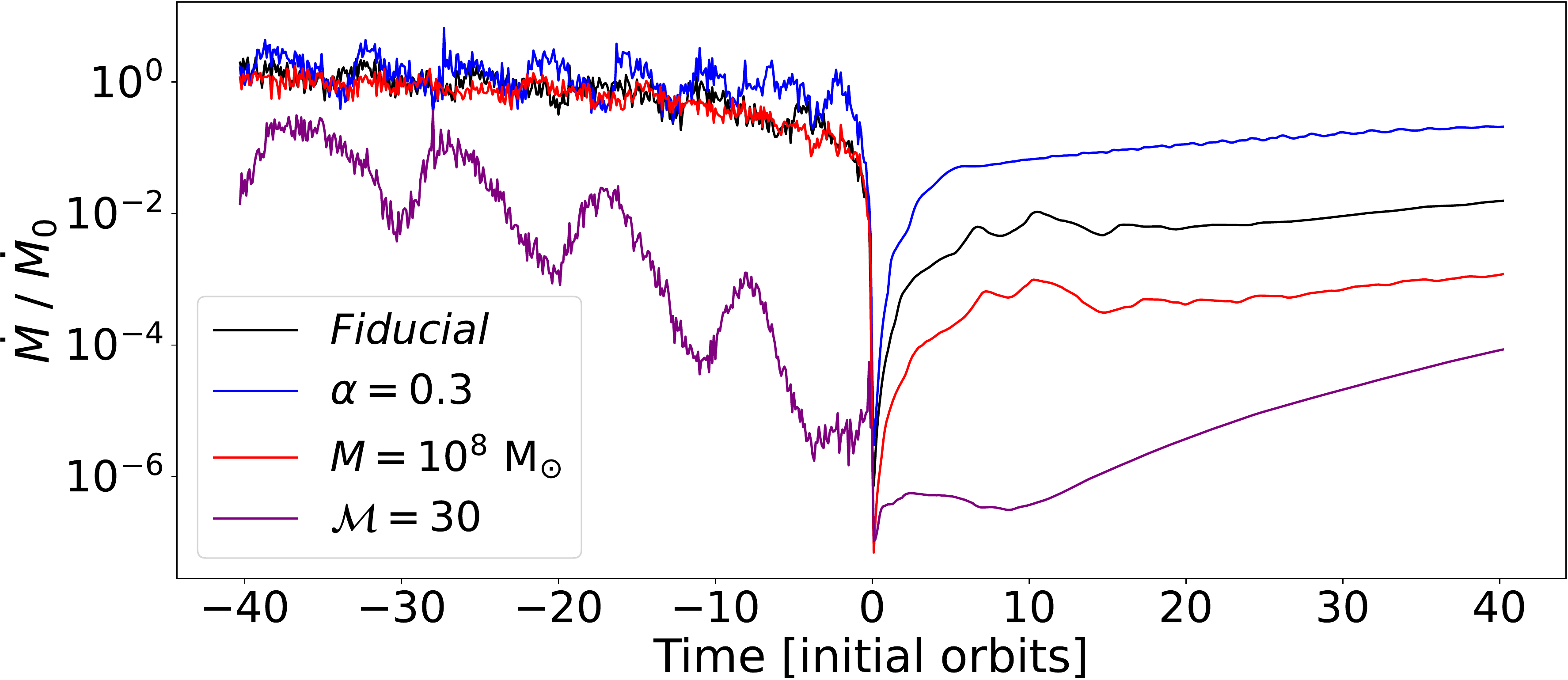}
    \caption{Accretion rates as a function of time (with time measured in units of the initial orbital time) for each model with no recoil or mass loss, normalised by their respective accretion rates for Shakura–Sunyaev discs around a single BH.}
    \label{fig:combined_accretions}
\end{figure}

\begin{figure}
    \centering
    \includegraphics[width=0.45\textwidth]{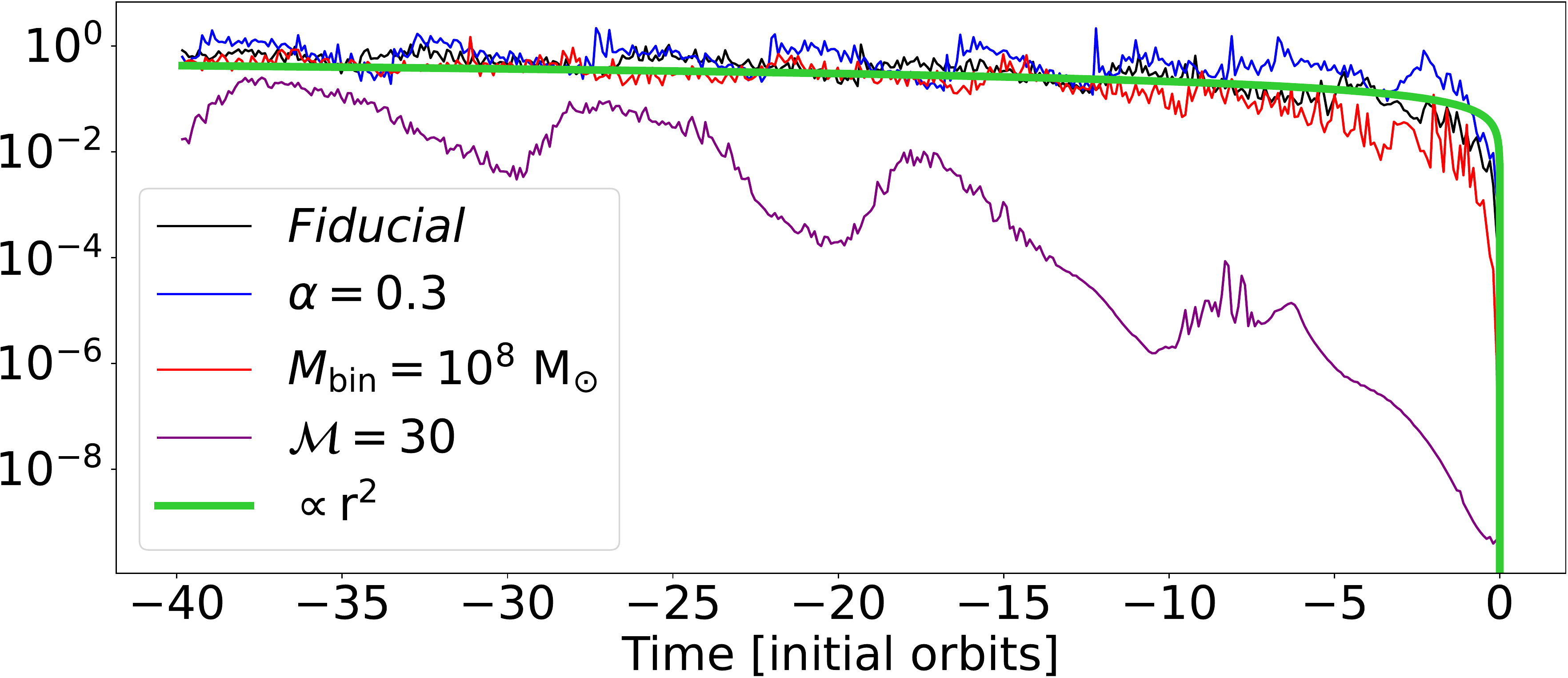}
    \caption{Normalised X-ray light curves in each model, and a (green) line proportional to the square of the binary separation $r^2$, representing the expected evolution of the surface area of tidally truncated  minidiscs as a function of time in initial orbits.}
    \label{fig:minidisc_size}
\end{figure}

To assess the robustness of our results we ran simulations with different values of several parameters, including: increasing viscosity to $\alpha=0.3$, increasing the total binary mass to $M_{\rm bin}=10^8$~\(\rm M_\odot\) and increasing the Mach number to $\mathcal{M} = 30$ (see Table~\ref{tab:Vars}). We also performed a resolution study, and tested if the disc remains geometrically thin during the full simulations, a necessary requirement for our vertically-integrated 2D approach. We present these numerical validation studies in the Appendix.

Figure~\ref{fig:bands_ratio_variations_combined} shows the fraction of the total luminosity emanating from the minidiscs for each model. Figure~\ref{fig:bands_variations_combined} shows the light curve in each band as labelled, and Figure~\ref{fig:combined_accretions} shows the accretion rates (normalised by their respective accretion rates for Shakura–Sunyaev discs around a single BH).

In the high-$\alpha$ model, the minidisc luminosity fractions are comparable to the fiducial model. The presence of large intermittent downward spikes is caused by hot material leaving the minidisc luminosity zone, which is somewhat arbitrarily defined. Again, we see a several-order-of-magnitude drop in the accretion rate (Fig.~\ref{fig:combined_accretions}) and the X-ray light curve (Fig.~\ref{fig:bands_variations_combined}) at the time of the merger, as in our fiducial model. The higher viscosity refills the post-merger cavity more rapidly, leading to faster recovery of the luminosity compared to our fiducial model. The lump in this model is more pronounced, leading to a larger amplitude lump periodicity in the light curve. The lump period is also shorter, likely due a smaller cavity caused by shorter viscous times. Additionally, the drop in UV luminosity is delayed; in the fiducial model, the decline is significant around 0.7 days prior to merger, whereas the decline is delayed until $\sim$0.3 days before merger in the $\alpha=0.3$ model. Again, this is likely because the shorter viscous time allows the cavity wall to follow the binary longer, keeping the minidiscs more full until their disruption just before merger.

In the high BH-mass model, not only is the X-ray emission almost exclusively from the minidiscs, as in the fiducial model, but the UV emission has increased to a much higher minidisc-fraction as well. This corresponds to a larger drop in the UV and bolometric luminosities near the time of merger. Again, we see a significant drop in the X-ray luminosity at merger, too, by a staggering 15 orders of magnitude. This is probably because there is more high-energy emission in a higher-mass system, and it is biased towards the hottest regions (the minidiscs).

In the higher Mach-number model, the minidiscs starve sooner during the inspiral. This can likely be attributed to the increased cavity size, specifically on the near-side of the cavity wall, making it even easier for the shrinking binary to outrun the cavity wall \citep[e.g.][]{ragusa2016}, and longer viscous time associated with the higher Mach number ($t_{\rm \nu}\sim \mathcal{M}^2$ if $\alpha$ and binary parameters are held fixed, as we did). This is our only simulation that begins inspiral past the nominal decoupling epoch, inspiralling over only 0.2 viscous times compared to the other models in which the inspiral lasts 1.7 (fiducial and $M_{\rm bin}=10^8$~\(\rm M_\odot\) cases), and 5.1 ($\alpha=0.3$ case) viscous times. This results in smaller minidisc fractions (Fig.~\ref{fig:bands_ratio_variations_combined}) and causes the drop in the X-ray flux to begin already several days before merger (Fig.~\ref{fig:bands_variations_combined}). However, the imprint of the longer lump period is more pronounced, which could help in the electromagnetic identification of these more massive binaries if a longer pre-merger temporal baseline is available. However, because the X-ray luminosity drops earlier, when LISA's sky localisation is still expected to be too poor to trigger EM monitoring~\citep{mangiagli2020}, one may need to rely on archival data to look for this flux decrease over several days prior to merger. Additionally, this is the only model in which the most luminous kick direction was less clear-cut; we show the "BH Kick Left" case, which may be the most luminous in the post-merger phase, but each kick direction had a similar luminosity at the end of our 5-day post-merger window.
The UV and bolometric luminosities are indistinguishable, since the UV band dominates, and they are remarkably steady across the merger. In fact, the small bump $\sim 0.75$ days after the merger only exists because of the kick imparted to the BH, otherwise, we find that these light curves remain flat. Together with our results from other models, this suggests that if an order of magnitude drop is observed in the UV near merger, then a much larger X-ray drop should accompany it, but if the UV is constant near merger, then the X-ray may already be below detectable levels.

Although it is too expensive with our current computational methods to evolve each of these models through the full recovery of the post-merger luminosity, in Figure~\ref{fig:combined_accretions} we show the mass accretion rates over the same time windows as in Figure~\ref{fig:bands_variations_combined}, but with time measured in units of the initial orbital time. The accretion rates are shown for no kick nor mass loss, as in Fig.~\ref{fig:fid_long_term_accretion}. In all but the Mach 30 case, we see a several-order-of-magnitude drop in the accretion rate at the merger, whereas in the Mach 30 model it decreases sooner, undulating downwards until the moment of merger, similar to the luminosity evolution in Fig.~\ref{fig:bands_variations_combined}.

Finally, in Figure~\ref{fig:minidisc_size}, we show each model's X-ray light curve, individually normalising the curves to their average value over one period about halfway through inspiral. As the binary shrinks, the tidal truncation radii of the minidiscs are reduced, scaling linearly with the orbital separation.  As a lowest-order picture, one may expect each minidisc's surface area to follow the binary separation $r$ as $\propto r^2$.  We show a curve representing this scaling, denoted in the legend as ``$\propto\!r^2$''. We see all but the Mach 30 light curves follow this $\propto r^2$ line, indicating that the rapid loss in luminosity can be attributed to the decreased surface area of the minidiscs. One expects that an enhanced drop below the $r^2$ line will occur to the extent that the rate of CBD feeding is not sufficient to keep the minidiscs saturated with material. This picture is consistent with the fact that the high-viscosity run ($\alpha=0.3$) stays above the $r^2$ line the most, and the fact that the high-Mach run ($\mathcal{M}=30$) drops below the $r^2$ line the most (the accretion rate is lowest in this run).  

\section{Observational prospects to identify electromagnetic counterparts of LISA sources}\label{subsec:xobs}

Identifying the host galaxy of merging MBHBs with networks of GW interferometers is difficult since they yield large sky localisation uncertainties. In the case of LISA, weeks before merger, the approximate position of the MBHB can be determined from the relative amplitudes and phases of the two polarisation components, from the periodic Doppler shift in the signal originating from the detector’s heliocentric motion, and from the additional modulation in the signal originating from the detector’s time-varying orientation~\citep{cutler1998}. However, it is not until near merger that the sky position can be significantly narrowed via additional effects - namely spin-precession, LISA's pattern response (which becomes frequency-dependent and assists with localisation;~\citealt{rubbo2004,marsat2021}), and higher-order harmonics (beyond the quadrupole; \citealt{baibhav2020,marsat2021}).

\citet{mangiagli2020} have recently studied sky localisation accuracy as a function of time prior to merger, using the most up-to-date LISA configuration and sensitivity curve, for a suite of MBHBs with varying total masses, mass ratios, spins, sky-positions and inclination angles, as well as different detector orientations. They find, in all cases, that the median error of sky-localisation decreases with time, but the dispersion around it increases. In a case near our fiducial model, they find that even as little as 10 hours before the merger, the typical uncertainties in the sky localisation are still near 10 deg$^2$, comparable to the field of view of large-scale optical/infrared survey telescopes, such as the Vera Rubin Observatory.
  
\citealt{lops2023a} have analysed the implication of these findings for a triggered campaign to find EM counterparts before merger. They considered multiple apparatus parameter configurations for both the Advanced Telescope for High-energy Astrophysics \citep[Athena;][]{nandra2013} and for the proposed Lynx mission \citep{thelynxteam2018}, and examined  binaries with varying total masses and redshifts. They assessed which binaries would be observable and identifiable 10 hours before, 1 hour before, and at merger via X-ray detection. For a binary near the mass of our fiducial model, at $z<1$, Lynx, with a FOV of 0.1 deg$^2$ and a spatial angular resolution of 0.5 arcseconds, can search the entire $\sim$10 deg$^2$ LISA error-box found by \citet{mangiagli2020} with $\sim$80 pointings, within 10 hours prior to merger. 
Although Athena actually has a larger FOV of 0.4 deg$^2$, it has a poorer angular resolution of 5-10 arcseconds and lower sensitivity, leading to higher exposure times required to detect the binaries. For our fiducial model of a $10^6~{\rm M_\odot}$ binary, assuming it shines near Eddington luminosity with $10-40\%$ of its bolometric flux in the X-ray band, within 10 hours prior to merger Athena can tile the LISA error-box with $\sim$5 pointings out to $z=0.5$ for hard X-rays in the 2-10 keV band, or with $\sim$20 pointings out to $z=1$ for soft X-rays in the 0.5-2 keV band. When looking ten hours before the merger, our results suggest emission may peak near 2 keV (see Figure~\ref{fig:fid_spectra_10hrsb4merger}) and decrease towards merger.

\citealt{canton2019} also explored tiling the LISA error box in the days leading to merger, with the proposed wide-field X-ray telescope Transient Astrophysics Probe (TAP), which has a FOV of 1 deg$^2$ and angular resolution of 5 arcseconds. While the focus of this paper was centred on detectability via the quasiperiodic Doppler modulation due to the orbital motion, their findings suggest a similar range of detectability, out to $z=1$. The probability of detection for our fiducial model decreases significantly beyond this redshift. More encouragingly, however, they find that a fraction of their simulated sources are not detected because of the number of photons collected is insufficient to measure the quasiperiodic modulations. As our method does not rely on this modulation, it should allow the detection of fainter sources to higher redshifts.

Previous proposals for identifying the host galaxy before merger require monitoring the pre-merger EM source for a prolonged time, in order to identify periodicity in the correct EM counterpart. By contrast, in principle, looking for the drop in thermal X-ray luminosity requires just two data points -- one just before or at merger, and one a few hours earlier -- because the counterpart will undergo a spectral change where the thermal X-rays disappear but stays relatively bright in the optical/UV. There is no reason for any other source among the many candidates in the LISA error volume to display this behaviour. This can significantly improve the chance of identifying the EM counterpart before merger occurs.

\section{Conclusions}
\label{sec:Con}

We performed 2D hydrodynamical simulations of inspiralling, accreting binaries, from the pre-decoupling epoch through the post-merger GW kick and mass loss, and examined the corresponding EM signatures. Compared to previous works, our study includes a longer inspiral period, a more physical viscosity, and a more realistic treatment of thermodynamics, directly solving the energy equation with a $\Gamma$-law equation of state for the gas, and with a physically-motivated cooling prescription. Additionally, compared to previous works \citep[e.g.][]{farris2015a,tang2018}, we use a much higher resolution which is important for understanding the late-stage minidisc destruction. From a suite of runs, we can draw the following conclusions:

\begin{enumerate}
    \item In all models prior to merger, the minidiscs surrounding the black holes emit essentially no luminosity in the optical band, but account for a sizeable fraction of the UV band, and nearly all of the X-ray band. 
    Most of the models in our suite, including the fiducial $M_{\rm bin}=10^6~\rm M_\odot$ binary, evolve across the so-called nominal decoupling time, i.e. when the GW inspiral outpaces the viscous timescale. 
    However, we see the minidiscs persist well past this nominal decoupling, often nearly until merger. 

    \item On the other hand, the minidiscs are eventually disrupted, which leads to a significant drop in both the accretion rates and the X-ray luminosity (and a smaller drop in the UV luminosity). In all models, this happens very near the time of merger, except in the Mach 30 case. Only in the Mach 30 model did we see any appreciable drop beginning earlier, with the accretion and thermal X-ray luminosity gradually decreasing until the merger, undulating along the way.

    \item Although BH recoil (or mass loss) can cause a non-negligible increase in post-merger luminosity through additional shocks, the plummet of the thermal X-ray luminosity still does not recover in the next several days post-merger. Mass loss tends to exacerbate the decrease in post-merger luminosity, extending the time needed to recover. 
   
    \item Sufficiently close to merger (within $\approx$10 hrs) and for certain binary masses and redshifts, Athena and Lynx could perform a full search of the LISA error box. Instead of requiring extensive pre-merger monitoring to extract a periodicity, our results suggest that as few as two data points are needed to identify the source via its disappearing thermal X-ray emission just before merger. 
    
    \item Accretion rates and light curves exhibit two periodicities, a multi-orbit one originating from the lump-modulation of accretion onto the minidiscs, and another on shorter timescales, arising from mass trading between minidiscs. Well before merger this latter period is close to the binary's orbital time, while at late stages, it is reduced to half of the orbital period.

Overall, our most important finding is that the minidiscs dominate the X-ray emission. The minidiscs last until moments before merger, but are ultimately disrupted. The corresponding sudden drop in the X-ray flux can provide a tell-tale EM signature, which can aid in the identification of the unique counterpart of the LISA GW source. Importantly, unlike attempts to identify pre-merger EM periodicity through extended monitoring, identifying an X-ray dropout is feasible with as few as two data points.

Our hydrodynamical models omit some physical ingredients that could alter the results we presented in this work. The most important ones that merit further investigation are three dimensional effects, magnetic fields, general relativity (GR), radiative feedback, and emission and feedback from jets and/or coronae. Both 3D \citep[e.g.][]{moody2019} and general-relativistic magnetohydrodynamics \citep[e.g.][]{combi2021,noble2021,avara2023} simulations indicate an increase in the variability of the lightcurve, although the fluctuation amplitudes (at most an order of magnitude) remain much smaller than the magnitude of the X-ray drop we predict here. GR would introduce several dynamical effects. A shallower post-Newtonian potential may also cause added variability and introduce an innermost stable circular orbit (ISCO), which can change the timing and magnitude of the X-ray drop in a way that also depends on the BH spins~\citep[e.g.][]{paschalidis2021,combi2022}. Ray-tracing, because of lensing near the BHs, results in weighting the portions of the minidisc that are closer to the BHs more heavily \citep[e.g.][]{davelaar2022}, again increasing variability, but unlikely to be able to hide a several-order-of magnitude drop in flux. Some 3D GRMHD findings \citep[e.g.][]{gutierrez2022} suggest that the thermal emission of the minidiscs may peak at a lower frequency than we find here, perhaps moving the large drop to the UV band. Radiative feedback \citep[e.g.][]{delvalle2018} could dismantle the minidiscs sooner, again hastening when the drop occurs and obscuring its coincidence with the merger. Additionally, as the minidiscs are disrupted, the black hole magnetosphere generates a split-monopole that will reconnect any remaining magnetic flux \citep[e.g.][]{bransgrove2021}. Depending on the reconnection time scales, this could generate a non-thermal X-ray component \citep[e.g.][]{hakobyan2023}, that could obscure the drop we observe in thermal X-ray emission. Interactions between misaligned jets may also lead to non-thermal emission at photon energies near (though perhaps not identical to) that of the minidiscs \citep[e.g.][]{gutierrez2023}. It is unclear if either dissipating jets or coronal contributions could obscure the drop in flux. More followup work will be needed to address these caveats.

As we were finalising this manuscript for submission, we became aware of a closely related preprint posted on the arXiv \citep{dittman2023}, addressing the gas dynamics of inspiralling binaries, similar to the present study. There are several notable differences in the methodology and emphasis. In particular, \citet{dittman2023} use a different grid code (\texttt{Athena++}), and perform a more extensive parameter study, including the dependence on viscosity, and show that decoupling can occur inside the LISA band for sufficiently high viscosities.  On the other hand, in addition to examining mass accretion rates, our study also includes predictions for the thermal emission, and includes the post-merger evolution.  Overall, the two studies find a consistent main result, namely that for comparable parameter choices (black hole masses and viscosities), the accretion rate drops by several orders of magnitude just prior to merger, providing a unique new EM signature of LISA mergers.

\end{enumerate}

\section*{Acknowledgements}

We acknowledge support by NSF grant AST-2006176 (ZH) and NASA grant 80NSSC22K082 (AM and ZH). JD acknowledges support by NASA grant NNX17AL82G and a Joint Columbia/Flatiron Postdoctoral Fellowship. Research at the Flatiron Institute is supported by the Simons Foundation. This research was supported in part by the National Science Foundation under Grant No. NSF PHY-1748958. This research has made use of NASA's Astrophysics Data System.
{\it Software:} {\tt python} \citep{travis2007,jarrod2011}, {\tt scipy} \citep{jones2001}, {\tt numpy} \citep{walt2011}, and {\tt matplotlib} \citep{hunter2007}.

%%%%%%%%%%%%%%%%%%%%%%%%%%%%%%%%%%%%%%%%%%%%%%%%%%
\section*{Data Availability}
The data underlying this article will be shared on reasonable request to the corresponding author.

\appendix

\section{Numerical validation}
\label{app-a}

\begin{figure}
    \centering
    \includegraphics[width=0.45\textwidth]{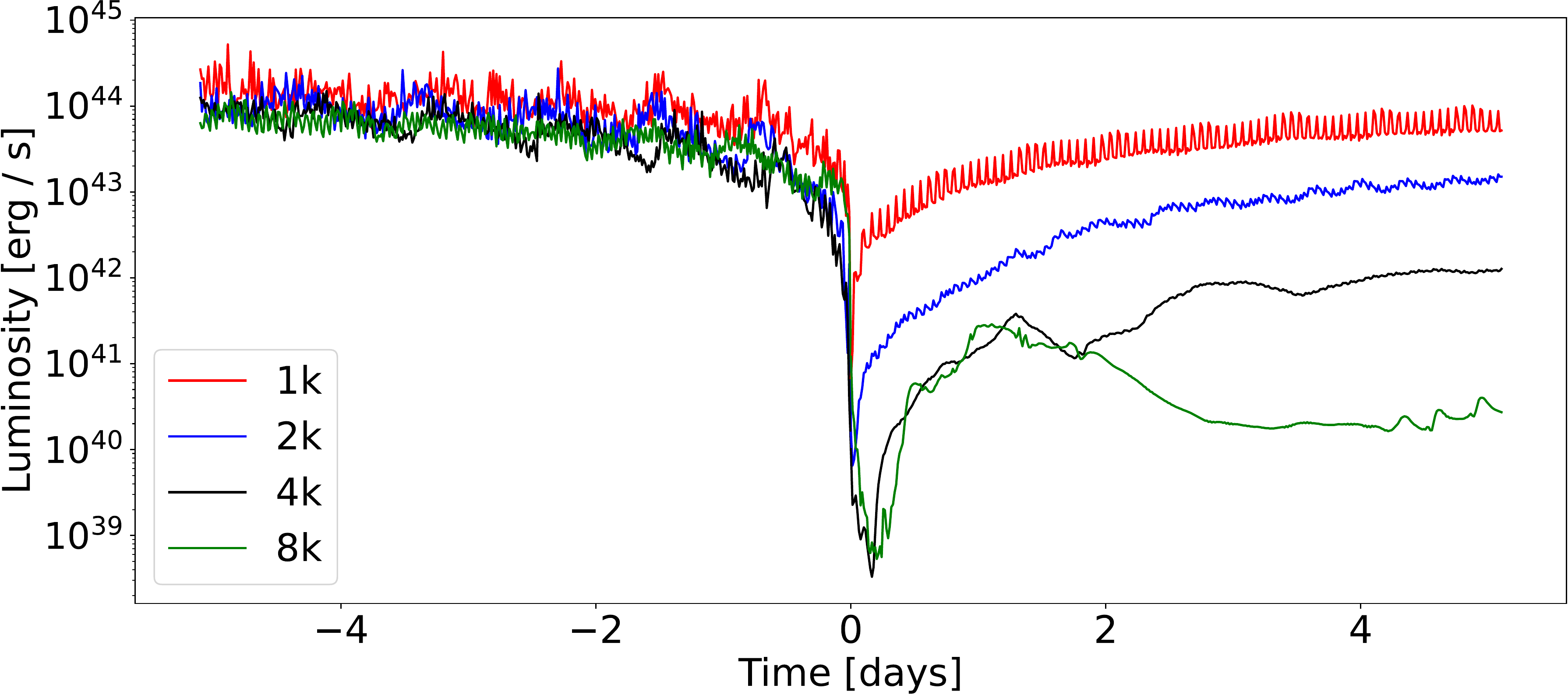}
    \caption{Resolution test of the X-ray luminosity for our fiducial model. The post-merger luminosity results from the downward BH with no mass-loss.}
    \label{fig:fid_res_1kto8k}
\end{figure}

To test our numerical convergence, we perform three variants of our fiducial run ($4000^2$ grid cells): two models with lower resolution ($1000^2$ and $2000^2$ grid cells), and one model with higher resolution ($8000^2$ grid cells). Figure~\ref{fig:fid_res_1kto8k} shows X-ray light curves, where 4k (black) indicates the fiducial model. The fiducial model appears robust near the window of merger we are most interested in. Note that verifying a converged result in this figure is complicated by the fact that the lump phase at merger is not equivalent across each run. Thus, the deviation between the 4k and 8k runs beginning two days after merger does not imply a lack of convergence at those times. Nonetheless, low resolution clearly over-predicts the post-merger luminosity significantly.

\begin{figure}

    \centering
    \includegraphics[width=0.45\textwidth]{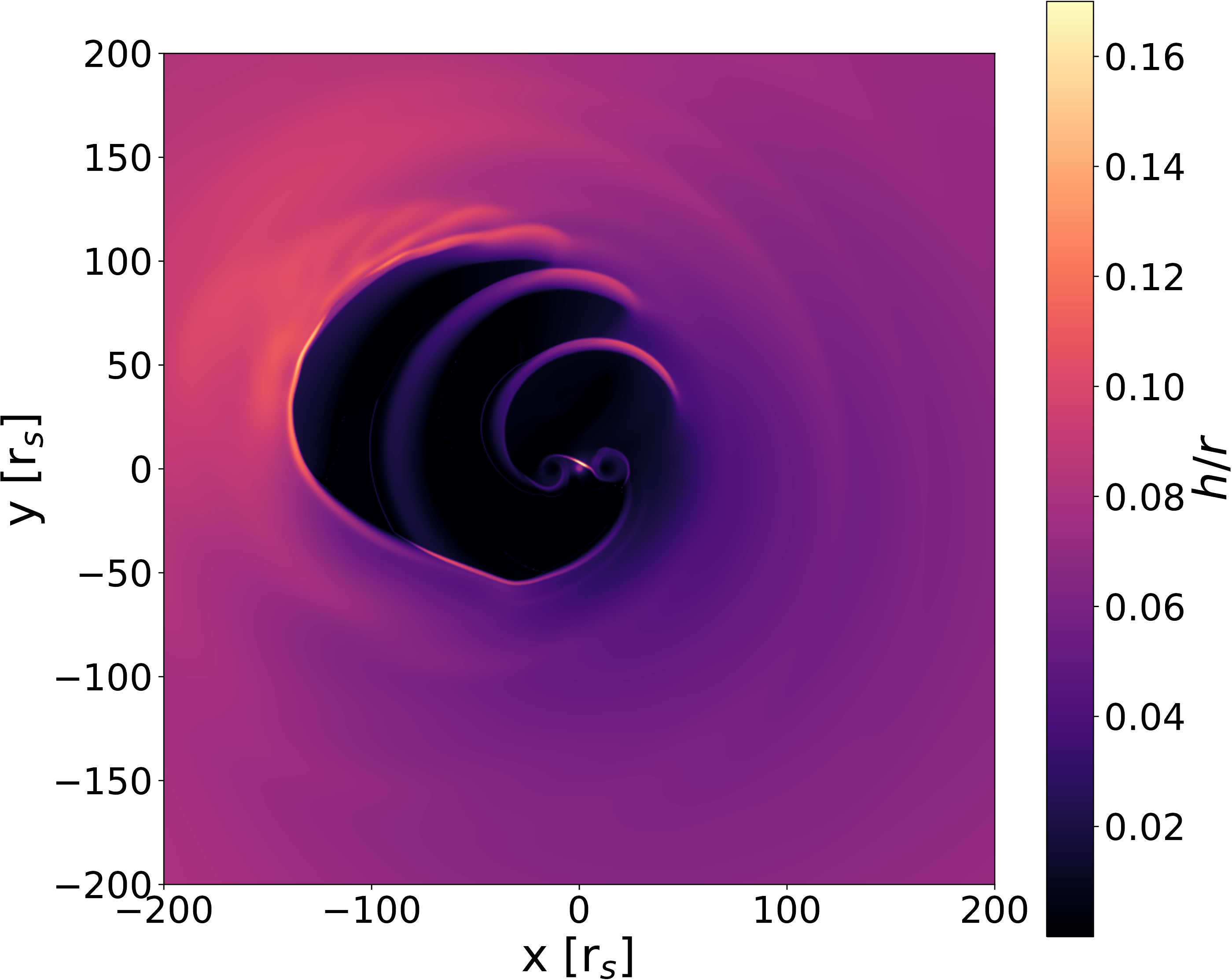}
    \caption{Map of the local disc thickness $h/r$ in the fiducial model, just before inspiral is initialised. }
    \label{fig:fid_premerger_hoverr}

    \centering
    \includegraphics[width=0.45\textwidth]{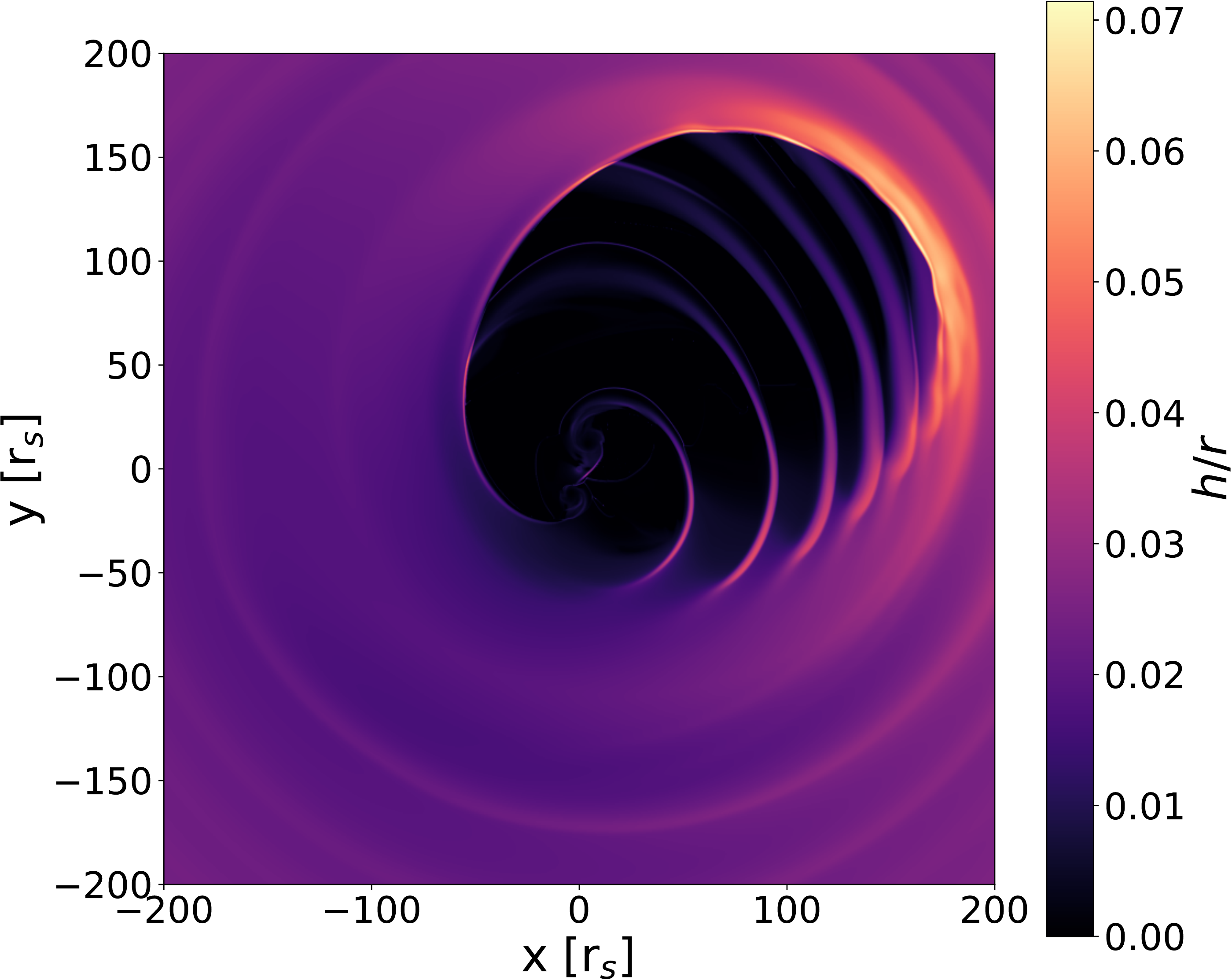}
    \caption{Map of the local disc thickness $h/r$ for the Mach=30 run, just before inspiral is initialised.}
    \label{fig:mach30_premerger_hoverr}
\end{figure}

Since our 2D column-integrated approach requires that the discs are geometrically thin, we checked the local aspect ratio $h/r \sim 1/\mathcal{M}$, shown in Figure~\ref{fig:fid_premerger_hoverr}. The overall $h/r$ is consistent with our intended setup, and the local disc aspect ratio remains valid. In the shock-heated regions, we see a rise in $h/r$, consistent with previous findings~\citep[e.g.][]{tang2018}; nevertheless, the disc remains thin, e.g. with at most $h/r\approx0.13$. We also checked the $\mathcal{M}=30$ simulation, shown in Figure~\ref{fig:mach30_premerger_hoverr}. We again see that the shock-heated regions show a rise in $h/r$, but the overall $h/r$ stays consistent with the thin disc assumption.

\section{Doppler effects}
\label{app-b}

Since this study was performed face-on, relativistic Doppler boosting was omitted, but given that $v \ll c$, this should be a minor effect. Here we consider results for inclined systems, for which Doppler effects are not negligible.

\begin{figure}
    \centering
    \includegraphics[width=0.45\textwidth]{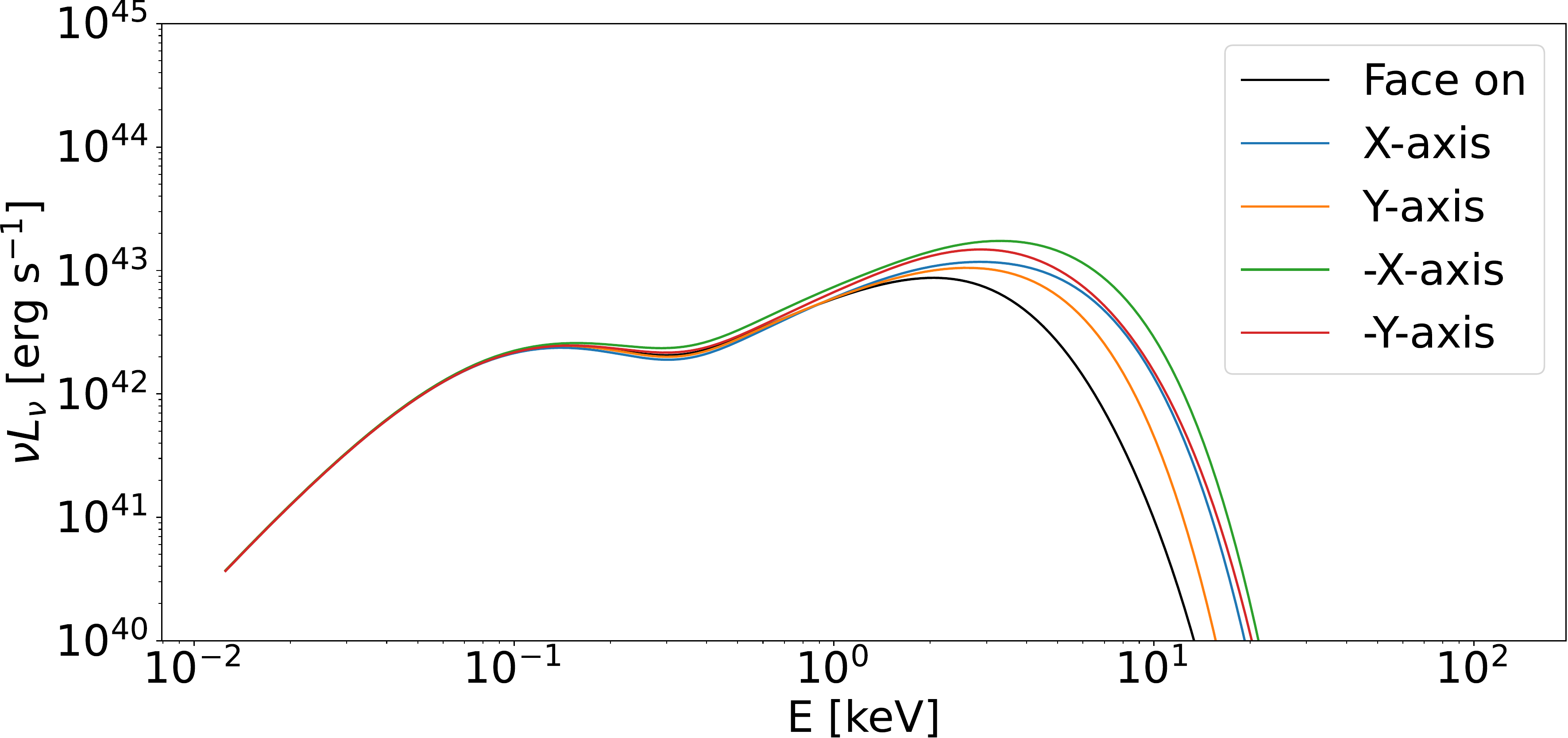}
    \caption{Spectra $\sim$10 hrs before merger from four different viewing angles - the X, Y, -X, and -Y axes -, incorporating relativistic Doppler effects. Independent of viewing angle, we see a brightening at higher frequencies, causing an overall hardening of the spectrum.}
    \label{fig:fid_spectra_10hrsb4merger}
\end{figure}

First, in Figure~\ref{fig:fid_spectra_10hrsb4merger}, we examine the spectra $\sim$10 hrs before merger for four different azimuthal viewing angles lying in the equatorial plane (an observer along the X, Y, -X, and -Y axes) to ascertain the upper limit of these effects. The spectra are computed in the same way as before, except that we include a relativistic Doppler boost from each patch of the disc. Practically, this is performed by modifying the effective temperature, following Eq. 3 of \citet{nakamura2009}. The figure shows that lower frequencies are unaffected, but the higher end of the spectrum tends to shift upwards in energy, independent of viewing angle. This is a result of the curvature in the Wien tail of the black-body spectrum, producing a larger brightening of the blue-shifted parts of the disc than the dimming of the patches red-shifted by a similar velocity~\citep[see also][]{tang2018}.

\begin{figure}
    \centering
    \includegraphics[width=0.45\textwidth]{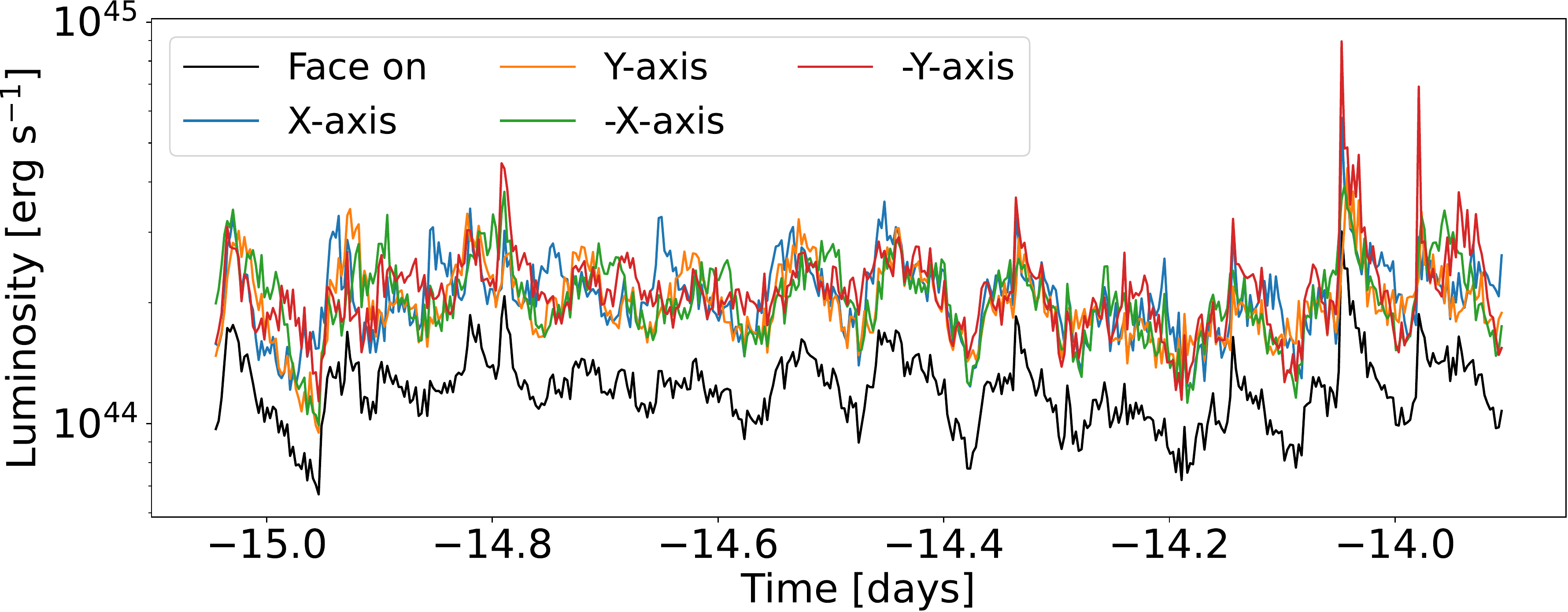}
    \caption{X-ray luminosity 15 to 14 days before merger, for viewing angles of face-on and along the X, Y, -X, and -Y axes.}
    \label{fig:fid_dopplerlc_15to14}

    \centering
    \includegraphics[width=0.45\textwidth]{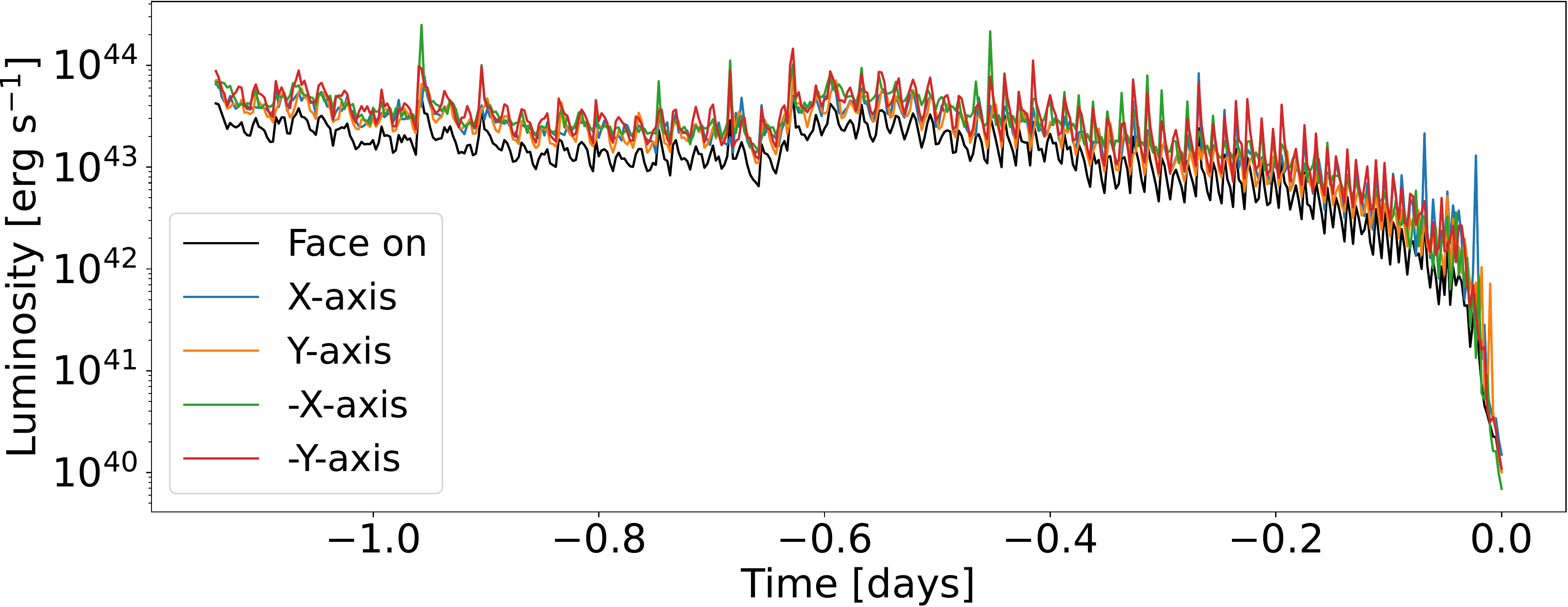}
    \caption{X-ray luminosity 1 to 0 days before merger, for viewing angles of face-on and along X, Y, -X, and -Y axes.}
    \label{fig:fid_dopplerlc_1to0}
\end{figure}

Next, in
Figures~\ref{fig:fid_dopplerlc_15to14} and~\ref{fig:fid_dopplerlc_1to0}, we examine the X-ray light curves from the same viewing angles, during the same time windows as in middle panels of Figure~\ref{fig:combined_strain_lum_pgram}. Corroborating what we see with the spectra, we see a boost to the overall luminosity in the X-ray band, independent of viewing angle. While there may be some increased variability, particularly at earlier times, we see a comparable signal and periodicity, and most importantly, we still see a significant drop in the X-ray luminosity at the time a merger.

Generically, the increased pre-merger luminosity due to the Doppler boost will help decrease the required exposure time for an X-ray triggered campaign, thereby increasing the mass and red-shift at which the entire LISA error-box can be tiled, as discussed in Section \S~\ref{subsec:xobs}. Specifically, in the case of Athena, however, because of a decreased sensitivity to harder X-rays, whether this would result in an overall increase or decrease to the exposure time will be system-dependent.

\label{lastpage}

\begin{thebibliography}{}
\makeatletter
\relax
\def\mn@urlcharsother{\let\do\@makeother \do\$\do\&\do\#\do\^\do\_\do\%\do\~}
\def\mn@doi{\begingroup\mn@urlcharsother \@ifnextchar [ {\mn@doi@}
  {\mn@doi@[]}}
\def\mn@doi@[#1]#2{\def\@tempa{#1}\ifx\@tempa\@empty \href
  {http://dx.doi.org/#2} {doi:#2}\else \href {http://dx.doi.org/#2} {#1}\fi
  \endgroup}
\def\mn@eprint#1#2{\mn@eprint@#1:#2::\@nil}
\def\mn@eprint@arXiv#1{\href {http://arxiv.org/abs/#1} {{\tt arXiv:#1}}}
\def\mn@eprint@dblp#1{\href {http://dblp.uni-trier.de/rec/bibtex/#1.xml}
  {dblp:#1}}
\def\mn@eprint@#1:#2:#3:#4\@nil{\def\@tempa {#1}\def\@tempb {#2}\def\@tempc
  {#3}\ifx \@tempc \@empty \let \@tempc \@tempb \let \@tempb \@tempa \fi \ifx
  \@tempb \@empty \def\@tempb {arXiv}\fi \@ifundefined
  {mn@eprint@\@tempb}{\@tempb:\@tempc}{\expandafter \expandafter \csname
  mn@eprint@\@tempb\endcsname \expandafter{\@tempc}}}

\bibitem[\protect\citeauthoryear{Acernese et~al.,}{Acernese
  et~al.}{2014}]{acernese2014}
Acernese F.,  et~al., 2014, \mn@doi [Classical and Quantum Gravity]
  {10.1088/0264-9381/32/2/024001}, 32, 024001

\bibitem[\protect\citeauthoryear{Akutsu et~al.,}{Akutsu
  et~al.}{2021}]{akutsu2021}
Akutsu T.,  et~al., 2021, \mn@doi [Progress of Theoretical and Experimental
  Physics] {10.1093/ptep/ptaa125}, 2021, 05A101

\bibitem[\protect\citeauthoryear{{Amaro-Seoane} et~al.,}{{Amaro-Seoane}
  et~al.}{2017}]{amaro-seoane2017}
{Amaro-Seoane} P.,  et~al., 2017, Laser {{Interferometer Space Antenna}}
  (\mn@eprint {arxiv} {arXiv:1702.00786}), \mn@doi{10.48550/arXiv.1702.00786}

\bibitem[\protect\citeauthoryear{Artymowicz \& Lubow}{Artymowicz \&
  Lubow}{1996}]{artymowicz1996}
Artymowicz P.,  Lubow S.~H.,  1996, \mn@doi [The Astrophysical Journal]
  {10.1086/310200}, 467, L77

\bibitem[\protect\citeauthoryear{{Avara}, {Krolik}, {Campanelli}, {Noble},
  {Bowen}  \& {Ryu}}{{Avara} et~al.}{2023}]{avara2023}
{Avara} M.~J.,  {Krolik} J.~H.,  {Campanelli} M.,  {Noble} S.~C.,  {Bowen} D.,
   {Ryu} T.,  2023, \mn@doi [arXiv e-prints] {10.48550/arXiv.2305.18538}, \href
  {https://ui.adsabs.harvard.edu/abs/2023arXiv230518538A} {p. arXiv:2305.18538}

\bibitem[\protect\citeauthoryear{Baibhav, Berti  \& Cardoso}{Baibhav
  et~al.}{2020}]{baibhav2020}
Baibhav V.,  Berti E.,   Cardoso V.,  2020, \mn@doi [Physical Review D]
  {10.1103/PhysRevD.101.084053}, 101, 084053

\bibitem[\protect\citeauthoryear{Baker, Boggs, Centrella, Kelly, McWilliams,
  Miller  \& {van Meter}}{Baker et~al.}{2008}]{baker2008}
Baker J.~G.,  Boggs W.~D.,  Centrella J.,  Kelly B.~J.,  McWilliams S.~T.,
  Miller M.~C.,   {van Meter} J.~R.,  2008, \mn@doi [The Astrophysical Journal]
  {10.1086/590927}, 682, L29

\bibitem[\protect\citeauthoryear{Baker et~al.,}{Baker et~al.}{2019}]{baker2019}
Baker J.,  et~al., 2019, Multimessenger Science Opportunities with {{mHz}}
  Gravitational Waves (\mn@eprint {arxiv} {arXiv:1903.04417}),
  \mn@doi{10.48550/arXiv.1903.04417}

\bibitem[\protect\citeauthoryear{Begelman, Blandford  \& Rees}{Begelman
  et~al.}{1980}]{begelman1980}
Begelman M.~C.,  Blandford R.~D.,   Rees M.~J.,  1980, \mn@doi [Nature]
  {10.1038/287307a0}, 287, 307

\bibitem[\protect\citeauthoryear{Bode, Haas, Bogdanovi{\'c}, Laguna  \&
  Shoemaker}{Bode et~al.}{2010}]{bode2010}
Bode T.,  Haas R.,  Bogdanovi{\'c} T.,  Laguna P.,   Shoemaker D.,  2010,
  \mn@doi [The Astrophysical Journal] {10.1088/0004-637X/715/2/1117}, 715, 1117

\bibitem[\protect\citeauthoryear{Bogdanovic, Miller  \& Blecha}{Bogdanovic
  et~al.}{2022}]{bogdanovic2022}
Bogdanovic T.,  Miller M.~C.,   Blecha L.,  2022, \mn@doi [Living Reviews in
  Relativity] {10.1007/s41114-022-00037-8}, 25, 3

\bibitem[\protect\citeauthoryear{Bowen, Campanelli, Krolik, Mewes  \&
  Noble}{Bowen et~al.}{2017}]{bowen2017}
Bowen D.~B.,  Campanelli M.,  Krolik J.~H.,  Mewes V.,   Noble S.~C.,  2017,
  \mn@doi [The Astrophysical Journal] {10.3847/1538-4357/aa63f3}, 838, 42

\bibitem[\protect\citeauthoryear{{Bransgrove}, {Ripperda}  \&
  {Philippov}}{{Bransgrove} et~al.}{2021}]{bransgrove2021}
{Bransgrove} A.,  {Ripperda} B.,   {Philippov} A.,  2021, \mn@doi [\prl]
  {10.1103/PhysRevLett.127.055101}, \href
  {https://ui.adsabs.harvard.edu/abs/2021PhRvL.127e5101B} {127, 055101}

\bibitem[\protect\citeauthoryear{Collaboration}{Collaboration}{2015}]{aLIGO2015}
Collaboration L.~S.,  2015, \mn@doi [Classical and Quantum Gravity]
  {10.1088/0264-9381/32/7/074001}, \href
  {https://ui.adsabs.harvard.edu/abs/2015CQGra..32g4001L} {32, 074001}

\bibitem[\protect\citeauthoryear{Combi, Armengol, Campanelli, Ireland, Noble,
  Nakano  \& Bowen}{Combi et~al.}{2021}]{combi2021}
Combi L.,  Armengol F. G.~L.,  Campanelli M.,  Ireland B.,  Noble S.~C.,
  Nakano H.,   Bowen .~D.,  2021, \mn@doi [Physical Review D]
  {10.1103/PhysRevD.104.044041}, 104, 044041

\bibitem[\protect\citeauthoryear{Combi, Armengol, Campanelli, Noble, Avara,
  Krolik  \& Bowen}{Combi et~al.}{2022}]{combi2022}
Combi L.,  Armengol F. G.~L.,  Campanelli M.,  Noble S.~C.,  Avara M.,  Krolik
  J.~H.,   Bowen D.~i.,  2022, \mn@doi [The Astrophysical Journal]
  {10.3847/1538-4357/ac532a}, 928, 187

\bibitem[\protect\citeauthoryear{Corrales, Haiman  \& MacFadyen}{Corrales
  et~al.}{2010}]{corrales2010}
Corrales L.~R.,  Haiman Z.,   MacFadyen A.,  2010, \mn@doi [Monthly Notices of
  the Royal Astronomical Society] {10.1111/j.1365-2966.2010.16324.x}, 404, 947

\bibitem[\protect\citeauthoryear{Cutler}{Cutler}{1998}]{cutler1998}
Cutler C.,  1998, \mn@doi [Physical Review D] {10.1103/PhysRevD.57.7089}, 57,
  7089

\bibitem[\protect\citeauthoryear{D'Orazio \& Di~Stefano}{D'Orazio \&
  Di~Stefano}{2018}]{dorazio2018}
D'Orazio D.~J.,  Di~Stefano R.,  2018, \mn@doi [Monthly Notices of the Royal
  Astronomical Society] {10.1093/mnras/stx2936}, 474, 2975

\bibitem[\protect\citeauthoryear{D'Orazio, Haiman  \& MacFadyen}{D'Orazio
  et~al.}{2013}]{dorazio2013}
D'Orazio D.~J.,  Haiman Z.,   MacFadyen A.,  2013, \mn@doi [Monthly Notices of
  the Royal Astronomical Society] {10.1093/mnras/stt1787}, 436, 2997

\bibitem[\protect\citeauthoryear{D'Orazio, Haiman  \& Schiminovich}{D'Orazio
  et~al.}{2015}]{dorazio2015}
D'Orazio D.~J.,  Haiman Z.,   Schiminovich D.,  2015, \mn@doi [Nature]
  {10.1038/nature15262}, 525, 351

\bibitem[\protect\citeauthoryear{{Dal Canton}, {Mangiagli}, {Noble},
  {Schnittman}, {Ptak}, {Klein}, {Sesana}  \& {Camp}}{{Dal Canton}
  et~al.}{2019}]{canton2019}
{Dal Canton} T.,  {Mangiagli} A.,  {Noble} S.~C.,  {Schnittman} J.,  {Ptak} A.,
   {Klein} A.,  {Sesana} A.,   {Camp} J.,  2019, \mn@doi [\apj]
  {10.3847/1538-4357/ab505a}, \href
  {https://ui.adsabs.harvard.edu/abs/2019ApJ...886..146D} {886, 146}

\bibitem[\protect\citeauthoryear{Davelaar \& Haiman}{Davelaar \&
  Haiman}{2022a}]{davelaar2022}
Davelaar J.,  Haiman Z.,  2022a, \mn@doi [Physical Review D]
  {10.1103/PhysRevD.105.103010}, 105, 103010

\bibitem[\protect\citeauthoryear{{Davelaar} \& {Haiman}}{{Davelaar} \&
  {Haiman}}{2022b}]{davelaar2022b}
{Davelaar} J.,  {Haiman} Z.,  2022b, \mn@doi [\prl]
  {10.1103/PhysRevLett.128.191101}, \href
  {https://ui.adsabs.harvard.edu/abs/2022PhRvL.128s1101D} {128, 191101}

\bibitem[\protect\citeauthoryear{Dempsey, Mu{\~n}oz  \& Lithwick}{Dempsey
  et~al.}{2020}]{dempsey2020}
Dempsey A.~M.,  Mu{\~n}oz D.,   Lithwick Y.,  2020, \mn@doi [The Astrophysical
  Journal] {10.3847/2041-8213/ab800e}, 892, L29

\bibitem[\protect\citeauthoryear{Dittmann \& Ryan}{Dittmann \&
  Ryan}{2021}]{dittmann2021}
Dittmann A.~J.,  Ryan G.,  2021, \mn@doi [The Astrophysical Journal]
  {10.3847/1538-4357/ac1bbd}, 921, 71

\bibitem[\protect\citeauthoryear{{Dittmann}, {Ryan}  \& {Miller}}{{Dittmann}
  et~al.}{2023}]{dittman2023}
{Dittmann} A.~J.,  {Ryan} G.,   {Miller} M.~C.,  2023, \mn@doi [arXiv e-prints]
  {10.48550/arXiv.2303.16204}, \href
  {https://ui.adsabs.harvard.edu/abs/2023arXiv230316204D} {p. arXiv:2303.16204}

\bibitem[\protect\citeauthoryear{Duffell, D'Orazio, Derdzinski, Haiman,
  MacFadyen, Rosen  \& Zrake}{Duffell et~al.}{2020}]{duffell2020}
Duffell P.~C.,  D'Orazio D.,  Derdzinski A.,  Haiman Z.,  MacFadyen A.,  Rosen
  A.~L.,   Zrake J.,  2020, \mn@doi [The Astrophysical Journal]
  {10.3847/1538-4357/abab95}, 901, 25

\bibitem[\protect\citeauthoryear{Farris, Duffell, MacFadyen  \& Haiman}{Farris
  et~al.}{2014}]{farris2014}
Farris B.~D.,  Duffell P.,  MacFadyen A.~I.,   Haiman Z.,  2014, \mn@doi [The
  Astrophysical Journal] {10.1088/0004-637X/783/2/134}, 783, 134

\bibitem[\protect\citeauthoryear{Farris, Duffell, MacFadyen  \& Haiman}{Farris
  et~al.}{2015a}]{farris2015}
Farris B.~D.,  Duffell P.,  MacFadyen A.~I.,   Haiman Z.,  2015a, \mn@doi
  [Monthly Notices of the Royal Astronomical Society: Letters]
  {10.1093/mnrasl/slu160}, 446, L36

\bibitem[\protect\citeauthoryear{Farris, Duffell, MacFadyen  \& Haiman}{Farris
  et~al.}{2015b}]{farris2015a}
Farris B.~D.,  Duffell P.,  MacFadyen A.~I.,   Haiman Z.,  2015b, \mn@doi
  [Monthly Notices of the Royal Astronomical Society: Letters]
  {10.1093/mnrasl/slu184}, 447, L80

\bibitem[\protect\citeauthoryear{Frank, King  \& Raine}{Frank
  et~al.}{2002}]{frank2002}
Frank J.,  King A.,   Raine D.~J.,  2002, Accretion {{Power}} in
  {{Astrophysics}}: {{Third Edition}}.
Cambridge University Press, \mn@doi{10.1017/CBO9781139164245}

\bibitem[\protect\citeauthoryear{Goodman}{Goodman}{2003}]{goodman2003}
Goodman J.,  2003, \mn@doi [Monthly Notices of the Royal Astronomical Society]
  {10.1046/j.1365-8711.2003.06241.x}, 339, 937

\bibitem[\protect\citeauthoryear{Guti{\'e}rrez, Combi, Noble, Campanelli,
  Krolik, Armengol  \& Garc{\'i}a}{Guti{\'e}rrez et~al.}{2022}]{gutierrez2022}
Guti{\'e}rrez E.~M.,  Combi L.,  Noble S.~C.,  Campanelli M.,  Krolik J.~H.,
  Armengol F.~L.,   Garc{\'i}a F.,  2022, \mn@doi [The Astrophysical Journal]
  {10.3847/1538-4357/ac56de}, 928, 137

\bibitem[\protect\citeauthoryear{Guti{\'e}rrez, Combi, Romero  \&
  Campanelli}{Guti{\'e}rrez et~al.}{2023}]{gutierrez2023}
Guti{\'e}rrez E.~M.,  Combi L.,  Romero G.~E.,   Campanelli M.,  2023, Flares
  from Dual Jets in Supermassive Black Hole Binaries (\mn@eprint {arxiv}
  {2301.04280}), \mn@doi{10.48550/arXiv.2301.04280}

\bibitem[\protect\citeauthoryear{Haiman}{Haiman}{2017}]{haiman2017}
Haiman Z.,  2017, \mn@doi [Physical Review D] {10.1103/PhysRevD.96.023004}, 96,
  023004

\bibitem[\protect\citeauthoryear{{Hakobyan}, {Ripperda}  \&
  {Philippov}}{{Hakobyan} et~al.}{2023}]{hakobyan2023}
{Hakobyan} H.,  {Ripperda} B.,   {Philippov} A.~A.,  2023, \mn@doi [\apjl]
  {10.3847/2041-8213/acb264}, \href
  {https://ui.adsabs.harvard.edu/abs/2023ApJ...943L..29H} {943, L29}

\bibitem[\protect\citeauthoryear{Hassan \& Rosen}{Hassan \&
  Rosen}{2012}]{hassan2012}
Hassan S.~F.,  Rosen R.~A.,  2012, \mn@doi [Physical Review Letters]
  {10.1103/PhysRevLett.108.041101}, 108, 041101

\bibitem[\protect\citeauthoryear{Hayasaki, Mineshige  \& Sudou}{Hayasaki
  et~al.}{2007}]{hayasaki2007}
Hayasaki K.,  Mineshige S.,   Sudou H.,  2007, \mn@doi [Publications of the
  Astronomical Society of Japan] {10.1093/pasj/59.2.427}, 59, 427

\bibitem[\protect\citeauthoryear{{Hunter}}{{Hunter}}{2007}]{hunter2007}
{Hunter} J.~D.,  2007, \mn@doi [Computing in Science and Engineering]
  {10.1109/MCSE.2007.55}, \href
  {https://ui.adsabs.harvard.edu/abs/2007CSE.....9...90H} {9, 90}

\bibitem[\protect\citeauthoryear{Jones, Oliphant, Peterson  et~al.}{Jones
  et~al.}{2001}]{jones2001}
Jones E.,  Oliphant T.,  Peterson P.,   et~al., 2001, {SciPy}: Open source
  scientific tools for {Python}, \url {http://www.scipy.org/}

\bibitem[\protect\citeauthoryear{{Kocsis}, {Haiman}  \& {Menou}}{{Kocsis}
  et~al.}{2008}]{kocsis2008}
{Kocsis} B.,  {Haiman} Z.,   {Menou} K.,  2008, \mn@doi [\apj]
  {10.1086/590230}, \href
  {https://ui.adsabs.harvard.edu/abs/2008ApJ...684..870K} {684, 870}

\bibitem[\protect\citeauthoryear{{Lightman} \& {Eardley}}{{Lightman} \&
  {Eardley}}{1974}]{lightman+1974}
{Lightman} A.~P.,  {Eardley} D.~M.,  1974, \mn@doi [\apjl] {10.1086/181377},
  \href {https://ui.adsabs.harvard.edu/abs/1974ApJ...187L...1L} {187, L1}

\bibitem[\protect\citeauthoryear{Lippai, Frei  \& Haiman}{Lippai
  et~al.}{2008}]{lippai2008}
Lippai Z.,  Frei Z.,   Haiman Z.,  2008, \mn@doi [The Astrophysical Journal]
  {10.1086/587034}, 676, L5

\bibitem[\protect\citeauthoryear{Liu, Wu  \& Cao}{Liu et~al.}{2003}]{liu2003}
Liu F.~K.,  Wu X.-B.,   Cao S.~L.,  2003, \mn@doi [Monthly Notices of the Royal
  Astronomical Society] {10.1046/j.1365-8711.2003.06235.x}, 340, 411

\bibitem[\protect\citeauthoryear{Lops, {Izquierdo-Villalba}, Colpi, Bonoli,
  Sesana  \& Mangiagli}{Lops et~al.}{2023}]{lops2023a}
Lops G.,  {Izquierdo-Villalba} D.,  Colpi M.,  Bonoli S.,  Sesana A.,
  Mangiagli A.,  2023, \mn@doi [Monthly Notices of the Royal Astronomical
  Society] {10.1093/mnras/stad058}, 519, 5962

\bibitem[\protect\citeauthoryear{Lousto \& Zlochower}{Lousto \&
  Zlochower}{2013}]{lousto2013}
Lousto C.~O.,  Zlochower Y.,  2013, \mn@doi [Physical Review D]
  {10.1103/PhysRevD.87.084027}, 87, 084027

\bibitem[\protect\citeauthoryear{MacFadyen \& Milosavljevic}{MacFadyen \&
  Milosavljevic}{2008}]{macfadyen2008}
MacFadyen A.~I.,  Milosavljevic M.,  2008, \mn@doi [The Astrophysical Journal]
  {10.1086/523869}, 672, 83

\bibitem[\protect\citeauthoryear{Mangiagli et~al.,}{Mangiagli
  et~al.}{2020}]{mangiagli2020}
Mangiagli A.,  et~al., 2020, \mn@doi [Physical Review D]
  {10.1103/PhysRevD.102.084056}, 102, 084056

\bibitem[\protect\citeauthoryear{Marsat, Baker  \& Canton}{Marsat
  et~al.}{2021}]{marsat2021}
Marsat S.,  Baker J.~G.,   Canton T.~D.,  2021, \mn@doi [Physical Review D]
  {10.1103/PhysRevD.103.083011}, 103, 083011

\bibitem[\protect\citeauthoryear{Megevand, Anderson, Frank, Hirschmann, Lehner,
  Liebling, Motl  \& Neilsen}{Megevand et~al.}{2009}]{megevand2009}
Megevand M.,  Anderson M.,  Frank J.,  Hirschmann E.~W.,  Lehner L.,  Liebling
  S.~L.,  Motl P.~M.,   Neilsen D.,  2009, \mn@doi [Physical Review D]
  {10.1103/PhysRevD.80.024012}, 80, 024012

\bibitem[\protect\citeauthoryear{{Mignon-Risse}, {Varniere}  \&
  {Casse}}{{Mignon-Risse} et~al.}{2023}]{mignon2023}
{Mignon-Risse} R.,  {Varniere} P.,   {Casse} F.,  2023, \mn@doi [\mnras]
  {10.1093/mnras/stac3794}, \href
  {https://ui.adsabs.harvard.edu/abs/2023MNRAS.519.2848M} {519, 2848}

\bibitem[\protect\citeauthoryear{Millman \& Aivazis}{Millman \&
  Aivazis}{2011}]{jarrod2011}
Millman K.~J.,  Aivazis M.,  2011, \mn@doi [Computing in Science \&
  Engineering] {10.1109/MCSE.2011.36}, 13, 9

\bibitem[\protect\citeauthoryear{Milosavljevi{\'c} \&
  Phinney}{Milosavljevi{\'c} \& Phinney}{2005}]{milosavljevic2005}
Milosavljevi{\'c} M.,  Phinney E.~S.,  2005, \mn@doi [The Astrophysical
  Journal] {10.1086/429618}, 622, L93

\bibitem[\protect\citeauthoryear{Moody, Shi  \& Stone}{Moody
  et~al.}{2019}]{moody2019}
Moody M. S.~L.,  Shi J.-M.,   Stone J.~M.,  2019, \mn@doi [The Astrophysical
  Journal] {10.3847/1538-4357/ab09ee}, 875, 66

\bibitem[\protect\citeauthoryear{{Nakamura}}{{Nakamura}}{2009}]{nakamura2009}
{Nakamura} T.~K.,  2009, \mn@doi [EPL (Europhysics Letters)]
  {10.1209/0295-5075/88/20004}, \href
  {https://ui.adsabs.harvard.edu/abs/2009EL.....8820004N} {88, 20004}

\bibitem[\protect\citeauthoryear{Nandra et~al.,}{Nandra
  et~al.}{2013}]{nandra2013}
Nandra K.,  et~al., 2013, The {{Hot}} and {{Energetic Universe}}: {{A White
  Paper}} Presenting the Science Theme Motivating the {{Athena}}+ Mission,
  \mn@doi{10.48550/arXiv.1306.2307}

\bibitem[\protect\citeauthoryear{Noble, Mundim, Nakano, Krolik, Campanelli,
  Zlochower  \& Yunes}{Noble et~al.}{2012}]{noble2012}
Noble S.~C.,  Mundim B.~C.,  Nakano H.,  Krolik J.~H.,  Campanelli M.,
  Zlochower Y.,   Yunes N.,  2012, \mn@doi [The Astrophysical Journal]
  {10.1088/0004-637X/755/1/51}, 755, 51

\bibitem[\protect\citeauthoryear{{Noble}, {Krolik}, {Campanelli}, {Zlochower},
  {Mundim}, {Nakano}  \& {Zilh{\~a}o}}{{Noble} et~al.}{2021}]{noble2021}
{Noble} S.~C.,  {Krolik} J.~H.,  {Campanelli} M.,  {Zlochower} Y.,  {Mundim}
  B.~C.,  {Nakano} H.,   {Zilh{\~a}o} M.,  2021, \mn@doi [\apj]
  {10.3847/1538-4357/ac2229}, \href
  {https://ui.adsabs.harvard.edu/abs/2021ApJ...922..175N} {922, 175}

\bibitem[\protect\citeauthoryear{O'Neill, Miller, Bogdanovi{\'c}, Reynolds  \&
  Schnittman}{O'Neill et~al.}{2009}]{oneill2009}
O'Neill S.~M.,  Miller M.~C.,  Bogdanovi{\'c} T.,  Reynolds C.~S.,   Schnittman
  J.~D.,  2009, \mn@doi [The Astrophysical Journal]
  {10.1088/0004-637X/700/1/859}, 700, 859

\bibitem[\protect\citeauthoryear{Oliphant}{Oliphant}{2007}]{travis2007}
Oliphant T.~E.,  2007, \mn@doi [Computing in Science \& Engineering]
  {10.1109/MCSE.2007.58}, 9, 10

\bibitem[\protect\citeauthoryear{Paczynski}{Paczynski}{1977}]{paczynski1977}
Paczynski B.,  1977, \mn@doi [The Astrophysical Journal] {10.1086/155526}, 216,
  822

\bibitem[\protect\citeauthoryear{{Paczy{\'n}sky} \& {Wiita}}{{Paczy{\'n}sky} \&
  {Wiita}}{1980}]{PW1980}
{Paczy{\'n}sky} B.,  {Wiita} P.~J.,  1980, \aap, \href
  {https://ui.adsabs.harvard.edu/abs/1980A&A....88...23P} {88, 23}

\bibitem[\protect\citeauthoryear{Paschalidis, Bright, Ruiz  \&
  Gold}{Paschalidis et~al.}{2021}]{paschalidis2021}
Paschalidis V.,  Bright J.,  Ruiz M.,   Gold R.,  2021, \mn@doi [The
  Astrophysical Journal Letters] {10.3847/2041-8213/abee21}, 910, L26

\bibitem[\protect\citeauthoryear{Penoyre \& Haiman}{Penoyre \&
  Haiman}{2018}]{penoyre2018}
Penoyre Z.,  Haiman Z.,  2018, \mn@doi [Monthly Notices of the Royal
  Astronomical Society] {10.1093/mnras/stx2469}, 473, 498

\bibitem[\protect\citeauthoryear{Peters}{Peters}{1964}]{Peters1964}
Peters P.~C.,  1964, \mn@doi [Physical Review] {10.1103/PhysRev.136.B1224},
  136, B1224

\bibitem[\protect\citeauthoryear{Pringle}{Pringle}{1991}]{pringle1991}
Pringle J.~E.,  1991, \mn@doi [Monthly Notices of the Royal Astronomical
  Society] {10.1093/mnras/248.4.754}, 248, 754

\bibitem[\protect\citeauthoryear{Ragusa, Lodato  \& Price}{Ragusa
  et~al.}{2016}]{ragusa2016}
Ragusa E.,  Lodato G.,   Price D.~J.,  2016, \mn@doi [Monthly Notices of the
  Royal Astronomical Society] {10.1093/mnras/stw1081}, 460, 1243

\bibitem[\protect\citeauthoryear{Roedig, Dotti, Sesana, Cuadra  \&
  Colpi}{Roedig et~al.}{2011}]{Roedig2011}
Roedig C.,  Dotti M.,  Sesana A.,  Cuadra J.,   Colpi M.,  2011, \mn@doi
  [Monthly Notices of the Royal Astronomical Society]
  {10.1111/j.1365-2966.2011.18927.x}, 415, 3033

\bibitem[\protect\citeauthoryear{{Rosotti}, {Lodato}  \& {Price}}{{Rosotti}
  et~al.}{2012}]{rosotti2012}
{Rosotti} G.~P.,  {Lodato} G.,   {Price} D.~J.,  2012, \mn@doi [\mnras]
  {10.1111/j.1365-2966.2012.21488.x}, \href
  {https://ui.adsabs.harvard.edu/abs/2012MNRAS.425.1958R} {425, 1958}

\bibitem[\protect\citeauthoryear{Rossi, Lodato, Armitage, Pringle  \&
  King}{Rossi et~al.}{2010}]{rossi2010}
Rossi E.~M.,  Lodato G.,  Armitage P.~J.,  Pringle J.~E.,   King A.~R.,  2010,
  \mn@doi [Monthly Notices of the Royal Astronomical Society]
  {10.1111/j.1365-2966.2009.15802.x}, 401, 2021

\bibitem[\protect\citeauthoryear{Rubbo, Cornish  \& Poujade}{Rubbo
  et~al.}{2004}]{rubbo2004}
Rubbo L.~J.,  Cornish N.~J.,   Poujade O.,  2004, \mn@doi [Physical Review D]
  {10.1103/PhysRevD.69.082003}, 69, 082003

\bibitem[\protect\citeauthoryear{Schnittman \& Krolik}{Schnittman \&
  Krolik}{2008}]{Schnittman2008}
Schnittman J.~D.,  Krolik J.~H.,  2008, \mn@doi [The Astrophysical Journal]
  {10.1086/590363}, 684, 835

\bibitem[\protect\citeauthoryear{Schutz}{Schutz}{1986}]{schutz1986}
Schutz B.~F.,  1986, \mn@doi [Nature] {10.1038/323310a0}, 323, 310

\bibitem[\protect\citeauthoryear{Shakura \& Sunyaev}{Shakura \&
  Sunyaev}{1973}]{shakura1973}
Shakura N.~I.,  Sunyaev R.~A.,  1973, \mn@doi [Astronomy and Astrophysics]
  {10.1017/S007418090010035X}, 24, 337

\bibitem[\protect\citeauthoryear{{Shakura} \& {Sunyaev}}{{Shakura} \&
  {Sunyaev}}{1976}]{shakura+1976}
{Shakura} N.~I.,  {Sunyaev} R.~A.,  1976, \mn@doi [\mnras]
  {10.1093/mnras/175.3.613}, \href
  {https://ui.adsabs.harvard.edu/abs/1976MNRAS.175..613S} {175, 613}

\bibitem[\protect\citeauthoryear{Shi \& Krolik}{Shi \& Krolik}{2015}]{shi2015}
Shi J.-M.,  Krolik J.~H.,  2015, \mn@doi [The Astrophysical Journal]
  {10.1088/0004-637X/807/2/131}, 807, 131

\bibitem[\protect\citeauthoryear{Shi, Krolik, Lubow  \& Hawley}{Shi
  et~al.}{2012}]{shi2012}
Shi J.-M.,  Krolik J.~H.,  Lubow S.~H.,   Hawley J.~F.,  2012, \mn@doi [The
  Astrophysical Journal] {10.1088/0004-637X/749/2/118}, 749, 118

\bibitem[\protect\citeauthoryear{Shields \& Bonning}{Shields \&
  Bonning}{2008}]{shields2008}
Shields G.~A.,  Bonning E.~W.,  2008, \mn@doi [The Astrophysical Journal]
  {10.1086/589427}, 682, 758

\bibitem[\protect\citeauthoryear{Tang, Haiman  \& MacFadyen}{Tang
  et~al.}{2018}]{tang2018}
Tang Y.,  Haiman Z.,   MacFadyen A.,  2018, \mn@doi [Monthly Notices of the
  Royal Astronomical Society] {10.1093/mnras/sty423}, 476, 2249

\bibitem[\protect\citeauthoryear{{The Lynx Team}}{{The Lynx
  Team}}{2018}]{thelynxteam2018}
{The Lynx Team} 2018, The {{Lynx Mission Concept Study Interim Report}},
  \mn@doi{10.48550/arXiv.1809.09642}

\bibitem[\protect\citeauthoryear{Tichy \& Marronetti}{Tichy \&
  Marronetti}{2008}]{tichy2008}
Tichy W.,  Marronetti P.,  2008, \mn@doi [Physical Review D]
  {10.1103/PhysRevD.78.081501}, 78, 081501

\bibitem[\protect\citeauthoryear{Vera C. Rubin~Collaboration et~al.,}{Vera C.
  Rubin~Collaboration et~al.}{2020}]{aLSST2020}
Vera C. Rubin~Collaboration V. C. R. O. L. S. S.~S.,  et~al., 2020, The
  {{Scientific Impact}} of the {{Vera C}}. {{Rubin Observatory}}'s {{Legacy
  Survey}} of {{Space}} and {{Time}} ({{LSST}}) for {{Solar System Science}}
  (\mn@eprint {arxiv} {arXiv:2009.07653}), \mn@doi{10.48550/arXiv.2009.07653}

\bibitem[\protect\citeauthoryear{{Westernacher-Schneider}, Zrake, MacFadyen  \&
  Haiman}{{Westernacher-Schneider} et~al.}{2022}]{westernacher-schneider2022}
{Westernacher-Schneider} J.~R.,  Zrake J.,  MacFadyen A.,   Haiman Z.,  2022,
  \mn@doi [Physical Review D] {10.1103/PhysRevD.106.103010}, 106, 103010

\bibitem[\protect\citeauthoryear{{Westernacher-Schneider}, {Zrake}, {MacFadyen}
   \& {Haiman}}{{Westernacher-Schneider}
  et~al.}{2023}]{westernacher-schneider2023}
{Westernacher-Schneider} J.~R.,  {Zrake} J.,  {MacFadyen} A.,   {Haiman} Z.,
  2023, \mn@doi [arXiv e-prints] {10.48550/arXiv.2307.01154}, \href
  {https://ui.adsabs.harvard.edu/abs/2023arXiv230701154W} {p. arXiv:2307.01154}

\bibitem[\protect\citeauthoryear{{Zanotti}, {Rezzolla}, {Del Zanna}  \&
  {Palenzuela}}{{Zanotti} et~al.}{2010}]{zanotti2010}
{Zanotti} O.,  {Rezzolla} L.,  {Del Zanna} L.,   {Palenzuela} C.,  2010,
  \mn@doi [\aap] {10.1051/0004-6361/201014969}, \href
  {https://ui.adsabs.harvard.edu/abs/2010A&A...523A...8Z} {523, A8}

\bibitem[\protect\citeauthoryear{{de Rham}, {Melville}  \& {Tolley}}{{de Rham}
  et~al.}{2018}]{rham2018}
{de Rham} C.,  {Melville} S.,   {Tolley} A.~J.,  2018, \mn@doi [Journal of High
  Energy Physics] {10.1007/JHEP04(2018)083}, \href
  {https://ui.adsabs.harvard.edu/abs/2018JHEP...04..083D} {2018, 83}

\bibitem[\protect\citeauthoryear{{del~Valle} \& Volonteri}{{del~Valle} \&
  Volonteri}{2018}]{delvalle2018}
{del~Valle} L.,  Volonteri M.,  2018, \mn@doi [Monthly Notices of the Royal
  Astronomical Society] {10.1093/mnras/sty1815}, 480, 439

\bibitem[\protect\citeauthoryear{{van der Walt}, {Colbert}  \&
  {Varoquaux}}{{van der Walt} et~al.}{2011}]{walt2011}
{van der Walt} S.,  {Colbert} S.~C.,   {Varoquaux} G.,  2011, \mn@doi
  [Computing in Science and Engineering] {10.1109/MCSE.2011.37}, \href
  {https://ui.adsabs.harvard.edu/abs/2011CSE....13b..22V} {13, 22}

\makeatother
\end{thebibliography}
\end{document}